\documentclass[10pt,journal,compsoc]{IEEEtran}

\usepackage{adjustbox}
\usepackage{afterpage}
\usepackage{balance}
\usepackage{booktabs}
\usepackage{caption}
\usepackage{cite}
\usepackage{colortbl}
\usepackage{collectbox}
\usepackage{comment}
\usepackage{enumitem}
\usepackage{framed}
\usepackage{graphicx}
\usepackage{lscape}
\usepackage{lipsum}
\usepackage{listings}
\usepackage{longtable}
\usepackage{multirow}
\usepackage{tabularx}
\usepackage{txfonts}
\usepackage{url}
\usepackage{hyperref}
\usepackage{varwidth}
\usepackage{xspace}
\usepackage[dvipsnames]{xcolor}
\usepackage[tight,footnotesize]{subfigure}
\usepackage{tcolorbox}
\tcbuselibrary{raster,skins}
\definecolor{light-gray}{gray}{0.85}

\newcommand{\countobservations}{
    \def \countobservations{1}
}
\newcounter{observation}
\newenvironment{observation}[1]
{   \refstepcounter{observation}
    \noindent \textbf{Observation~\theobservation) \textit{#1}}
}

\countobservations

\newcommand{\countimplications}{
    \def \countimplications{1}
}
\newcounter{implication}
\newenvironment{implication}[1]
{
    \refstepcounter{implication}
    \noindent \textbf{Implication~\theimplication ) \textit{#1}}
}

\countimplications

\newcommand{\countmess}{
    \def \countmess{1}
}
\newcounter{mes}
\countmess

\def\summarybox#1#2{
\medskip
\begin{tcolorbox}[enhanced,title=#1, attach boxed title to top left= {xshift=0mm, yshift*=-\tcboxedtitleheight/2}]
    #2
\end{tcolorbox}
}

\newlist{stepitemize}{itemize}{1}
\setlist[stepitemize,1]{leftmargin=1.26cm}

\definecolor{diffstart}{named}{Gray}
\definecolor{diffincl}{named}{Green}
\definecolor{diffrem}{named}{YellowOrange}
\lstdefinelanguage{diff}{
basicstyle=\ttfamily\small,
float,
floatplacement=t,
frame=tb,
morecomment=[f][\color{diffstart}]{@@},
morecomment=[f][\color{diffincl}]{+},
morecomment=[f][\color{diffrem}]{-},
}

\newcommand{\RQone}{How many proposals get declined or subsequently resubmitted?}
\newcommand{\RQtwo}{Can declined proposals be anticipated?} 
\newcommand{\RQthree}{What characterizes declined proposals?}

\def\et{et\ al.\xspace}
\def\ie{i.e.\xspace}

\def\fig#1{Figure~\ref{#1}}

\def\tab#1{Table~\ref{#1}}

\def\sec#1{Section~\ref{#1}}

\newlength\MAX  

\usepackage{framed}
\usepackage{xparse}
\newsavebox{\fminipagebox}
\NewDocumentEnvironment{hassanbox}{m O{\fboxsep}}
 {\par\kern#2\noindent\begin{lrbox}{\fminipagebox}
  \begin{minipage}{#1}\ignorespaces}
 {\end{minipage}\end{lrbox}%
  \makebox[#1]{%
    \kern\dimexpr-\fboxsep-\fboxrule\relax
    \fbox{\usebox{\fminipagebox}}%
    \kern\dimexpr-\fboxsep-\fboxrule\relax
  }\par\kern#2
 }

\title{An empirical study on declined proposals:\\why are these proposals declined?}

\author{\IEEEauthorblockN{Masanari Kondo,}
\and
\IEEEauthorblockN{Mahmoud Alfadel}
\and
\IEEEauthorblockN{Shane McIntosh}
\and
\IEEEauthorblockN{Yasutaka Kamei,}
\and
\IEEEauthorblockN{Naoyasu Ubayashi,}
}

\author{Masanari Kondo,
        Mahmoud Alfadel,
        Shane McIntosh,
        Naoyasu Ubayashi,
        Yasutaka Kamei,
\IEEEcompsocitemizethanks{
\IEEEcompsocthanksitem
Masanari Kondo, and Yasutaka Kamei are with the
Graduate School and Faulty of Information Science and Electrical Engineering,
Kyushu University, Japan.\protect\\
E-mail: {kondo,kamei}@ait.kyushu-u.ac.jp
\IEEEcompsocthanksitem
Mahmoud Alfadel is with the Department of Computer Science,
University of Calgary, Canada
\protect\\
E-mail: mahmoud.alfadel@ucalgary.ca
\IEEEcompsocthanksitem
Shane McIntosh is with the David R. Cheriton School of Computer Science,
University of Waterloo, Canada.
\protect\\
E-mail: shane.mcintosh@uwaterloo.ca
\IEEEcompsocthanksitem
Naoyasu Ubayashi is with the Faculty of Science and Engineering, Waseda University, Japan
\protect\\
E-mail: ubayashi@aoni.waseda.jp
}% <-this % stops an unwanted space
\thanks{Manuscript received XX YY, 2020; revised XX XX, 2021.}}

\IEEEtitleabstractindextext{%
\begin{abstract}
\textbf{Background:}
Design-level decisions in open-source software (OSS) projects are often made through structured mechanisms such as \emph{proposals}, which require substantial community discussion and review. 
Despite their importance, the proposal process is resource-intensive and often leads to contributor frustration, especially when proposals are declined without clear feedback. 
Yet, the reasons behind proposal rejection remain poorly understood, limiting opportunities to streamline the process or guide contributors effectively. 

\textbf{Objective:}
This study investigates the characteristics and outcomes of proposals in the Go programming language to understand why proposals are declined and how such outcomes might be anticipated. 

\textbf{Methods:}
We conduct a mixed-method empirical study on 1,091 proposals submitted to the Go project. We quantify proposal outcomes, build a taxonomy of decline reasons, and evaluate large language models (LLMs) for predicting these outcomes. 

\textbf{Results:}
We find that proposals are more often declined than accepted, and resolution typically takes over a month. Only 14.7\% of declined proposals are ever resubmitted. Through qualitative coding, we identify nine key reasons for proposal decline, such as duplication, limited use cases, or violations of project principles. This taxonomy can help contributors address issues in advance, e.g., checking for existing alternatives can reduce redundancy.
We also demonstrate that GPT-based models can predict decline decisions early in the discussion (F1 score = 0.71 with partial comments), offering a practical tool for prioritizing review effort.

\textbf{Conclusion:}
Our findings reveal inefficiencies in the proposal process and highlight actionable opportunities for improving both contributor experience and reviewer workload by enabling early triage and guiding contributors to strengthen their proposals using a structured understanding of past decline reasons.
\end{abstract}

}

\begin{document}

\maketitle

\IEEEdisplaynontitleabstractindextext

\IEEEpeerreviewmaketitle

\section{Introduction}
Open-source software (OSS) maintainers receive various \emph{modification requests (MR)} from contributors~\cite{pinto2016SANER,gousios2015ICSE}. These MRs are discussed, and some are accepted while others are declined.
This decision-making processes, along with the reasoning behind each decision known as \emph{rationales}~\cite{zannier2007IST}, can be time-consuming for maintainers~\cite{li2022TSE}.
MRs cover a wide range of decisions from code-level decisions to design-level decisions. Code-level decisions accompany the changes in the source code such as formatting. In contrast, design-level decisions are more abstract such as the introduction of a new language feature (e.g., generics in Go~\cite{go-generics}).

The design-level decisions are more important for maintainers than other decisions since once such a design-level MR is accepted, it will significantly change the software and maintainers have to maintain it for a long time period. Hence, maintainers need to carefully consider design-level decisions.
If maintainers could anticipate whether a design-level MR would be accepted or declined, along with understanding the reasons, it would save considerable effort for contributors.
Maintainers could leverage this information to streamline their review process, focusing their efforts on MRs with a higher chance of acceptance, while also providing clearer justifications for declining others.
Hence, we believe that it is important to investigate design-level decisions in OSS projects and develop a prediction model to anticipate whether a design-level MR will be accepted or declined.

However, design-level decisions are less documented compared to other decisions such as code-level ones and are less investigated in software engineering research. For example, pull requests, which consist of code-level decisions, have been extensively studied~\cite{li2022TSE,gousios2014ICSE,tsay2014ICSE,iyer2021TSE,rastogi2016ICSE,rastogi2018ESEM,kononenko2018ICSE,pinto2018CHASE,terrell2017,steinmacher2018ICSE,weissgerber2008MSR,jiang2013MSR,baysal2013WCRE,lenarduzzi2021JSS,rahman2016ICSE,soares2015ICMLA,soares2015,rahman2014MSR,zhang2023TSE}. Li~\et~\cite{li2022TSE} empirically investigated the abandonment of pull requests and identified 12 reasons. Dey and Mockus proposed a prediction model to determine the acceptance likelihood of pull requests~\cite{dey2020ESEM}.

In this paper, we examine the characteristics of design-level decisions and propose a method to anticipate whether a design-level MR will be accepted or declined.
For our study, we focus on \emph{proposals} in the Go language~\cite{go}, which represent MRs in the Go development process.
This process is particularly strict compared to other projects, with a well-documented and transparent design-level decision-making procedure.
We believe that this rigorous process results in high-quality data, enabling us to thoroughly analyze the characteristics of design-level decisions (\ie, declined proposals) and develop an accurate prediction method.

Since anticipating whether a proposal will be accepted or declined is a binary classification problem, we mainly focused on a declined proposal. We developed a prediction model to anticipate whether a proposal will be declined.
We examined the Go proposals and addressed the research questions (RQs) below.

\begin{itemize}
  \item[\textbf{(RQ1)}] \textbf{\RQone{}}
  \\ 
  \emph{Motivation:}
  To get an initial understanding of the declined proposals, we analyzed their quantitative characteristics.
  \\
  \emph{Results:}
  More than 40\% of the proposals were declined, with only 14.7\% were resubmitted. Therefore, a considerable number of proposals are abandoned.
  \item[\textbf{(RQ2)}] \textbf{\RQthree{}}
  \\
  \emph{Motivation:}
  To deeply grasp why these proposals were declined, we used an open coding approach to create a taxonomy of the reasons.
  \\
  \emph{Results:}
  We identified nine reasons for declined proposals, all of which are related to a comprehensive understanding of Go. Anticipating whether a proposal will be declined or accepted before submission requires contributors to have in-depth knowledge of Go.
  \item[\textbf{(RQ3)}] \textbf{\RQtwo{}}
  \\ 
  \emph{Motivation:}
  If maintainers and contributors could identify declined proposals without the discussion, it would reduce the workload for both maintainers and contributors.
  Hence, we develop a prediction method to determine if a proposal will be declined.
  \\
  \emph{Results:}
  GPT-3.5-turbo can effectively anticipate declined proposals, achieving an F1 score of 0.88. However, without considering the discussion within the proposal, the F1 score drops to 0.67.
  % \item[\textbf{(RQ4)}] \textbf{\RQfour{}}
  % \\
  % \emph{Motivation:}
  % Similar to RQ2, if maintainers and contributors could anticipate the reasons for declined proposals, it would reduce the workload for both maintainers and contributors. Hence, we investigate whether we can anticipate the reasons for declined proposals. 
  % \\
  % \emph{Results:}
  % GPT-3.5-turbo can accurately anticipate Duplication and Existing Alternatives, achieving over 60\% accuracy. However, the accuracy for other reasons is less than 40\%. GPT-4o improves the performance for the subjective reasons but not for the other reasons.
\end{itemize}

Our contributions are as follows:
\begin{itemize}
  \item We conducted a large-scale empirical study on the design-level decisions and quantified the characteristics of declined proposals in Go development. Moreover, we manually analyzed declined proposals and identified nine declined reasons.
  \item We developed a prediction model to anticipate whether a proposal will be declined, achieving an F1 score of 0.88. 
  %Moreover, we also developed a prediction model to anticipate the declined reasons, achieving over 60\% accuracy for Duplication and Existing Alternatives.
  \item We provided future research directions to improve the performance of the prediction model. 
  \item We made a publicly available replication package to facilitate future research on design-level decisions.\footnote{We will replace the following temporary link with the official Zenodo DOI link once this paper is accepted: \url{https://www.dropbox.com/scl/fo/45txk4gix8nmv6qqlryno/AGAe5hk9uwi34_jnxzD70Vc?rlkey=qrk1la4d0ykrdbhcpq4olvq3t&st=tjcohzpm&dl=0}}
\end{itemize}

% \noindent{}
% \textbf{Paper Organization.}
% \sec{sec:back} explains the proposals in Go development.
% \sec{sec:design} describes the study design. 
% \sec{sec:rq1}, \ref{sec:rq2}, and \ref{sec:rq3} present the results of RQ1, RQ2, and RQ3, respectively.
% \sec{sec:discussion} discusses the implications of our observations. 
% \sec{sec:rw} compares our work with related studies.
% \sec{sec:threats} provides the threats to validity. 
% \sec{sec:conclusion} concludes the paper and explains the recommendations.

\section{Proposals in Go Community}
\label{sec:back}
The Go community follows a workflow for discussing and adjudicating new features suggested by contributors.\footnote{\url{https://go.googlesource.com/proposal/+/master/README.md}} This workflow ensures transparent decision-making and collaboration.

In this section, we describe the proposal workflow in the Go community (Section~\ref{sec:back:proposal}) and present an example to explain why we focus our study on declined proposals (Section~\ref{sec:back:example}).
% Finally, we review related studies to contextualize our research study (Section~\ref{sec:back:related}).

\subsection{Proposal Workflow}
\label{sec:back:proposal}
%In this research project, we focus on the software development history of the Go programming language.
% The Go community follows a strict rule for discussing MRs and making decisions, known as proposals~\cite{go}.
% The Go community follows a structured process for discussing and deciding on new features. 
% This process, referred to as \textit{the proposal process},\footnote{\url{https://go.googlesource.com/proposal/+/master/README.md}} ensures transparent decision-making and collaboration.
The workflow consists of the following steps:

\begin{enumerate}
  \item
A contributor who wishes to propose a new feature, creates an issue labeled as a proposal.
  \item
%Maintainers add an active label, and contributors discuss the proposal.
Maintainers and contributors discuss the proposal.
  \item
%Based on the discussion, maintainers decide whether to add a likely accept label or a likely decline label, and contributors continue discussing the proposal.
Based on the discussion, maintainers classify the proposal as ``likely to be accepted'' or ``likely to be declined''.
  \item
%Finally, maintainers make a decision by either adding an accepted label or a declined label.
Maintainers make a final decision, marking the proposal as either ``accepted'' or ``declined''.
\end{enumerate}

%\mahmoud{For this para, I feel we are repeating ourselves. In fact, the next subsection makes the concept of Proposal more clear. Hence, I'd suggest removing this para from this subsection and integrate it with the motivating example. Hence, Fig 1 might not be needed.}

%% ============================================================================
%% PLEASE DON"T REMOVE THIS COMMENT
%% ============================================================================
%%
%% ============================================================================
%% ============================================================================	

\begin{figure}[t]
  \centering
      \includegraphics[width=0.9\columnwidth]{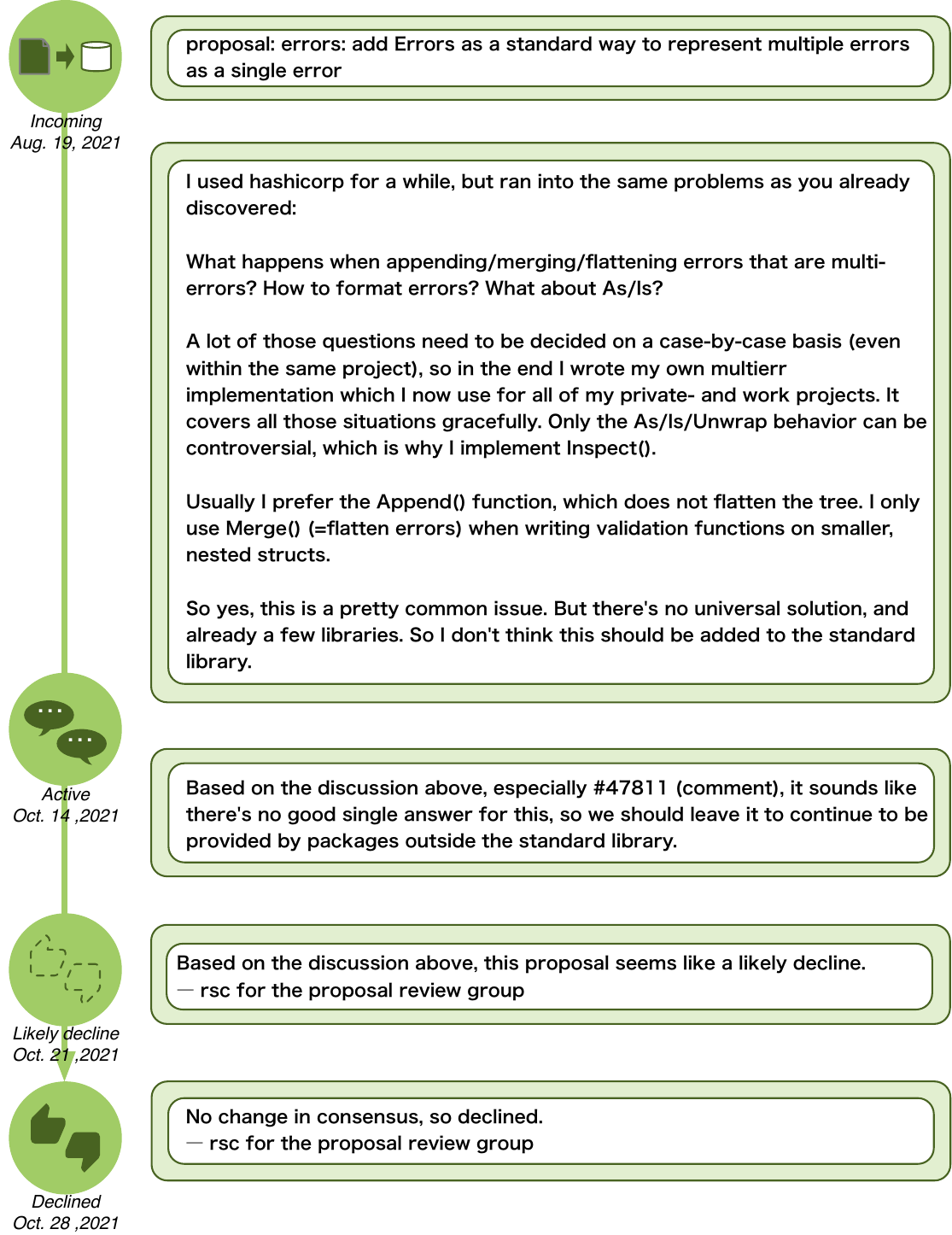}
  \caption{An example declined proposal retrieved from \texttt{issues/47811} in \texttt{golang/go}~\cite{go-issue-ex}}
  \label{fig:background:example}
\end{figure}
\subsection{Motivating Example}
\label{sec:back:example}
% This section presents an example to illustrate the motivation behind our study of proposals in the Go community and provides a detailed explanation of the different proposal statuses.

Proposals move between \emph{incoming}, \emph{active}, \emph{likely-accept}, \emph{likely-decline}, \emph{accepted}, and \emph{declined} states.
\fig{fig:background:example} illustrates these states using a concrete example. When a proposal is submitted, it enters the \emph{incoming} state, indicating that it is awaiting initial triage. 
This example proposes adding a standard way to handle multiple errors in Go's \texttt{errors} package, enabling developers to combine them into a single error for easier management and reporting.
Once submitted, the proposal transitions to the \emph{active} state, where maintainers and contributors engage in discussions.
After this initial discussion, maintainers decide whether to classify the proposal as \emph{likely-accept} or \emph{likely-decline}. 
In this case, a maintainer initially marked the proposal as \emph{likely-decline} and ultimately \emph{declined} it.

At first glance, the proposal workflow may seem similar to that of Pull Requests (PRs), a feature of social coding platform, such as GitHub, which is used in collaborative software development. PRs allow developers to propose changes to a codebase, which are then peer reviewed before being merged.
% Prior research has extensively examined the factors influencing PR rejections~\cite{li2022TSE,gousios2014ICSE,tsay2014ICSE,iyer2021TSE,rastogi2016ICSE,rastogi2018ESEM,kononenko2018ICSE,pinto2018CHASE,terrell2017,steinmacher2018ICSE,weissgerber2008MSR,jiang2013MSR,baysal2013WCRE,lenarduzzi2021JSS,rahman2016ICSE,soares2015ICMLA,soares2015,rahman2014MSR,zhang2023TSE}.

Proposals differ from PRs in their scope. While PRs focus on code-level changes, proposals involve \emph{design-level decisions} that influence the language and its broader ecosystem. For example, the declined proposal discussed earlier suggested a new approach to error handling in Go---an architectural change affecting the language as a whole rather than a specific code contribution.

Declined proposals often arise from architectural considerations rather than code quality or correctness, making it essential to understand their underlying factors. Our study examines the reasons behind declined proposals and develops approaches to inferring declined proposals and their causes.

\begin{figure}[t]
  \centering
  \includegraphics[width=\columnwidth]{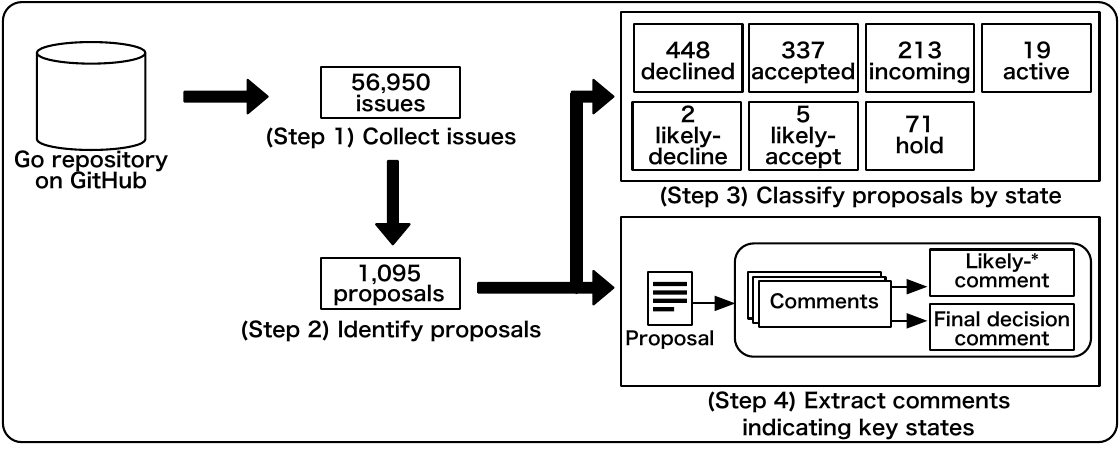}
  \caption{Overview of data collection steps and RQs}
  \label{fig:data-collection}
\end{figure}

\section{Study Design}
\label{sec:design}
% I renamed the section name to make it more abstract and reflect the bigger picture of the study. Also, I'd suggest we move Section 2.1 (from Background section) to be subsection in this section; it can be used as a motivation/justification of studying Go community.
% 1. GitHubのGoのissuesをlabels，projectCards（incomimngとかstatusが保存されている），そのcommentsと共に取り出す
% 2. LabelsにProposalという名前があるissuesを取り出す
% 3. 2で分類したうちprojectCardsにdeclinedとあるもの（448件）を分析対象とする．RQ1ではacceptedなproposals（337件）も分析対象である.
% In this section, we describe our rationale for examining proposals in Go (\sec{sec:community}) and explain our approaches to data extraction (\sec{sec:data-ext}).
Proposals in Go address design-level decisions and are discussed and adjudicated based on stringent criteria, making them a rich source of data for our analysis.
Therefore, to investigate the characteristics of declined design-level MRs, we focus our study on proposals from the Go language repository.
\fig{fig:data-collection} provides an overview of the steps we follow for our data collection and analysis. 
Each step is described in detail below.

% \begin{figure}[tb!]
%   \centering
%   \includegraphics[width=\columnwidth]{figures/data/data-collection.pdf}
%   \caption{Overview of the steps for our data collection and analysis.}\mahmoud{Let us change the word "stage" to "status", since we use status in previous sections. Also, can you please check the matching between each step's title here and in the section, since I made some changes there, just a consistency issue.}
%   \label{fig:data-collection}
% \end{figure}

%% ============================================================================
%% PLEASE DON"T REMOVE THIS COMMENT
%% ============================================================================
%% exp3/05_statistics.ipynb
%% ============================================================================
%% ============================================================================
\subsection{Collect issues}
\label{sec:design:collect-issues}
Proposals in Go are a specific type of GitHub issue used to suggest and discuss substantial design-level changes or enhancements to the language, its core libraries, or its toolchain.
Since our study aims to examine proposals submitted to Go, we begin by extracting all issues from the Go project repository.\footnote{\label{fn:go}\url{https://github.com/golang/go}} 
We use the GitHub API\footnote{\url{https://docs.github.com/en/graphql}} to retrieve all issues submitted to the Go repository. 
This query yields a total of 56,950 issues, as collected in September 2023.
%\masa{The community uses V2 before this date so this does not affect the data. Should we mention this?}

\subsection{Identify proposals}
% \textbf{(Step 2) Identifying proposals.}
Proposals in Go are managed through the \textit{Proposals} project board.\footnote{\url{https://github.com/orgs/golang/projects/17/views/1}}
This board organizes GitHub issues labeled as ``Proposal'' by their state in the proposal workflow. At any point in time, each proposal state is incoming, active, likely-decline, likely-accept, declined, accepted, and hold.
For example, issue \#45861, titled ``\textit{proposal: doc: make an official go-standards/project-layout documentation within Go project}'' is labeled as a proposal and included in the project board.
As of the time of this writing, the issue is ``active'' and appears in the board that are also in that state. 

There are Classic and V2 types of proposals.
Classic is a previous version of the proposals while V2 is the current version.
Due to restrictions on accessing V2 proposals through the GitHub API\footnote{As of the time of collecting the data, Go prohibits access to and download V2 proposals through the GitHub API.}, our analysis is limited to Classic proposals.
%\masa{Need to discuss. I think the classic proposal was closed and we cannot access it. They also changed the name...}
To identify Classic proposals, we use the GitHub API to extract issues labeled as ``Proposal''.
Upon applying this step, we identify 1,095 Classic proposals.
Because all of these proposals were submitted before September 2023, collecting them in September 2023 (the issue collection process in \sec{sec:design:collect-issues}) does not influence our analysis.

% \vspace{1mm}
% \noindent{}
\subsection{Classify proposals by state}
\label{step3}
% \textbf{(Step 3) Classifying proposals by stage:}
%~\mahmoud{I added the word 'by stage' because I feel that in this step we do not really classify but we just identify another type of info, which is the stage.}
To examine the rates of accepted and declined proposals, we need to determine the state of each collected proposal. 
To achieve this, we inspect the \texttt{projectCards} attribute, which provides detailed information about the current states (e.g., accepted, likely-accepted, incoming) of each proposal.
% Since proposals have several statuses (\eg, accepted, likely-accepted, and incoming), we classified them accordingly using the projectCards attribute.
% \fig{fig:proposal-statuses} shows the number of proposals for each status.~\mahmoud{I feel we should not add this figure here since this looks more related to results in RQ1. We can only describe the process of identifying the stage of each proposal.}
% To understand the entire process from incoming status to accepted or declined status, we investigated the accepted and declined proposals in this paper.

% \vspace{1mm}
% \noindent{}
\subsection{Extract comments indicating key states}
%\masa{My idea is to explain the details why using likely decisions affect the prediction model. What do you think?}
Our study contributes to the development of a prediction model that anticipates the final decision of a proposal and the reasoning behind it, based on comments made throughout its lifecycle.  
To prevent \textit{data leakage}--where the model accesses information unavailable at the time of prediction--this step identifies comments associated with an \emph{initial likely decision} (e.g., ``likely-accept'') and excludes these comments and any subsequent ones from the input data.  
This ensures that the model relies only on information available before this stage when making predictions.

%\subsubsection{Identifying likely-decision comments}
The workflow for proposals has two stages. First, maintainers assess whether a proposal is likely to be accepted or declined, and this preliminary decision is often documented in a comment (\fig{fig:background:example}).  
Before such comments are submitted, contributors (i.e., developers who submitted proposals) and maintainers usually engage in extensive discussions to refine the proposal.
Second, after the initial likely decision, maintainers finalize and officially announce the decision. In most cases, the final decision aligns with the initial likely decision.

To identify initial likely decision comments, we apply {keyword-based matching} to detect explicit indications of the likely outcome of a proposal. 
We use six predefined keywords: \emph{accept}, \emph{declin} (used instead of ``decline'' to match variations like ``declining''), \emph{likely accept}, \emph{likely-accept}, \emph{likely declin}, \emph{likely-declin}.
%\masa{I have no idea how do we justify these keywords.}

We focus on the 785 proposals of the 1,095 collected proposals where a final decision had been made--{448 declined} and {337 accepted} proposals. There are 17,665 comments associated with the 785 proposals in our dataset.  
A keyword search across these comments identified the {735 proposals} containing at least one relevant keyword. We exclude the 50 proposals that did not match any of the keywords, since our inspection did not reveal any explicit decision-related comments.

We note that there are proposals where the likely stage in the workflow is skipped and the proposal proceeds directly to the final decision (i.e., the identified keywords were either \emph{accept} or \emph{decline}).  
To account for such cases, we inspect the 735 proposals to identify instances where the likely stage had been skipped, observing 104 of such proposals. In these cases, we consider the final decision comment (\ie, containing \emph{accept} or \emph{decline}) as the initial likely decision comment. 
This step does not impact our data analysis. Regardless of the number of such cases, as long as these comments in the identified proposal are excluded from the prediction data, our approach remains unaffected.

\section{\RQone{}}
%\section{Quantitative Analysis}
\label{sec:rq1}
%RQ1: \emph{\RQone{}}
%\subsection{Motivation}
%\mahmoud{I feel the reader would prefer to have the Motivation subsection separated from the Approach subsection?}
% However, not all proposals are accepted, raising questions about its effectiveness. A low decline rate may suggest that the workflow effectively guides contributors toward successful submissions, while a high decline rate could indicate inefficiencies, such as unclear guidelines or systemic biases against certain types of changes.
%The Go proposal workflow serves as a structured mechanism for introducing changes to the language.
The Go proposal workflow is the process by which changes are made to the language.
%This workflow guides maintainers toward making successful decisions on proposals.
Understanding decision trends, such as decline rates, may provide insight into the decision-making process and its impact on the evolution of the language.

Proposals that are initially declined may be refined and resubmitted as new proposals.
Frequent resubmissions would suggest that contributors adapt proposals based on feedback to align with acceptance criteria, whereas rare resubmissions may suggest that resubmissions are at least tacitly discouraged, which may lead to missed opportunities.
Investigating the rate at which declined proposals in Go are resubmitted will help to assess whether the proposal workflow fosters refinement and iteration, or whether declined ideas tend to be entirely abandoned.

\begin{figure}[t]
  \centering
      \includegraphics[width=1.0\columnwidth]{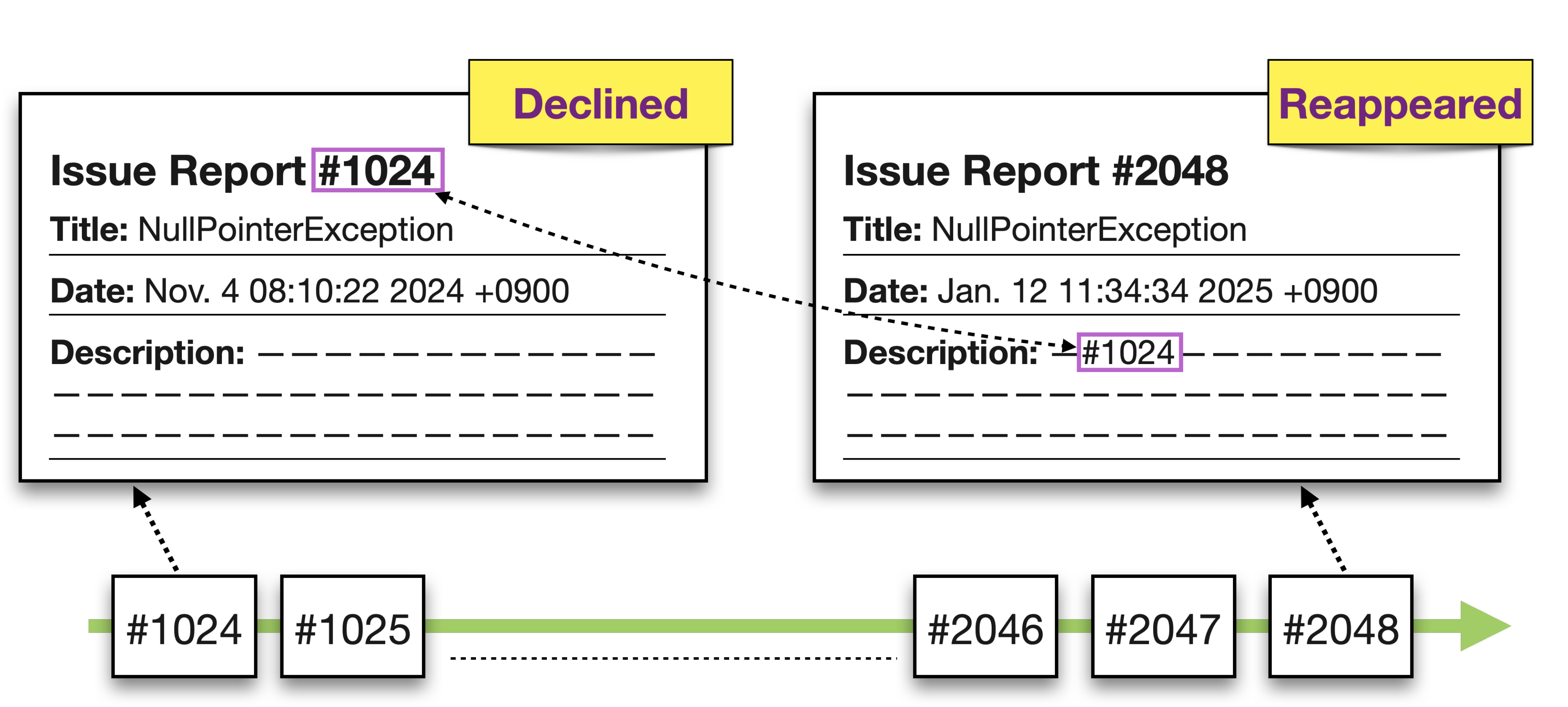}
  \caption{An example of the keyword approach}
  \label{fig:ex-keyword-approach}
\end{figure}

\subsection{Approach}
%To explore the two aspects above, we analyze (1) the frequency of declined proposals and (2) how often declined proposals were resubmitted as new proposals.
%Proposals follow the workflow and can have different states, such as accepted, declined, and hold. We retrieve this information using the GitHub API.
To address RQ1, we first determine the state of each proposal in our dataset using the procedure from \sec{sec:design}. 
Once the states have been identified, we compute the frequency of each state to determine the proportion of proposals that are declined relative to other outcomes.
Additionally, we examine their resolution times to assess the duration required to reach an initial decision.

To investigate resubmitted proposals, we use GitHub issue numbers to establish links between declined proposals and other related issues. 
Specifically, we use a keyword approach~\cite{kondo2022EMSE}, which treats issue numbers as references.
\fig{fig:ex-keyword-approach} shows an example of this approach. Issue \texttt{\#1024} is a declined proposal, and issue \texttt{\#2048} is another proposal. The description of \texttt{\#2048} includes the issue number of \texttt{\#1024}, indicating that \texttt{\#2048} is a resubmitted proposal associated with \texttt{\#1024}. Once we identify a resubmitted proposal, we label the associated declined proposal has resubmitted.
To identify resubmitted proposals, the keyword approach searches for issue numbers of declined proposals in the descriptions of other issues.
%This approach allows us to link issues to software artifacts (\eg, commit messages) if they contain the issue numbers as references.
%We apply this approach to connect declined proposals with other issues. If the description of an issue includes the issue number of a declined proposal, we consider that declined proposal to have reappeared as this issue.\mahmoud{I tried to rephrase this but I still feel it is hard to understand the approach here. I'd suggest we bring a clear and real example and illustrate the approach using the example.}

%\input{figures/rq1/num_issues_category.tex}
%% ============================================================================
%% PLEASE DON"T REMOVE THIS COMMENT
%% ============================================================================
%% replication/rq1/06_make_table.ipynb
%% replication/rq1/tables/rq1_num_issue_category.csv
%% ============================================================================
%% ============================================================================	

\begin{table}[t]
  \caption{Proportion of proposal states}
  \label{tab:rq1:proposal-statuses}
\begin{center}
\scalebox{1.0}{
    \begin{tabular}{lrr}
    \toprule
    \multicolumn{1}{c}{\textbf{State}}&\multicolumn{1}{c}{\textbf{\#Proposals}}&\multicolumn{1}{c}{\textbf{\%Proposals}}\\
    \midrule
    Declined&448&40.9\\
    Accepted&337&30.8\\
    Incoming&213&19.5\\
    Hold&71&6.5\\
    Active&19&1.7\\
    Likely-accept&5&0.5\\
    Likely-decline&2&0.2\\
    \bottomrule
    \end{tabular}
    }
    \end{center}
\end{table}

%\input{tables/rq1/linked-proposals.tex}
%\tab{tab:rq1:linked-proposals}
%\input{figures/rq1/all_linked_issues.tex}
%\input{figures/rq1/reappeared_issue_results.tex}
%% ============================================================================
%% PLEASE DON"T REMOVE THIS COMMENT
%% ============================================================================
%% replication/rq1/06_make_table.ipynb
%% replication/rq1/tables/rq1_resubmitted.csv
%% ============================================================================
%% ============================================================================	

% \begin{table}[t]
\begin{table}[t]
    \caption{Proportion of accepted/declined resubmitted proposals}
\label{tab:rq1:reappeared-proposal-results}
\begin{center}
\scalebox{1.0}{
    \begin{tabular}{lrr}
    \toprule
    \multicolumn{1}{c}{\textbf{State}}&\multicolumn{1}{c}{\textbf{\#Proposals}}&\multicolumn{1}{c}{\textbf{\%Proposals}}\\
    %\midrule
    %Resubmitted&66&\multicolumn{1}{c}{--}\\
    \midrule
    Declined&40&53.3\\
    Accept&19&25.3\\
    WIP&16&21.3\\
    \bottomrule
    \end{tabular}
    }
    \end{center}
\end{table}

%% ============================================================================
%% PLEASE DON"T REMOVE THIS COMMENT
%% ============================================================================
%% exp20/figures/proposal-resolution-days.pdf
%% exp20/01_01_resolution_time.ipynb
%% ============================================================================
%% ============================================================================	

\begin{figure}[t]
  \centering
      \includegraphics[width=\columnwidth]{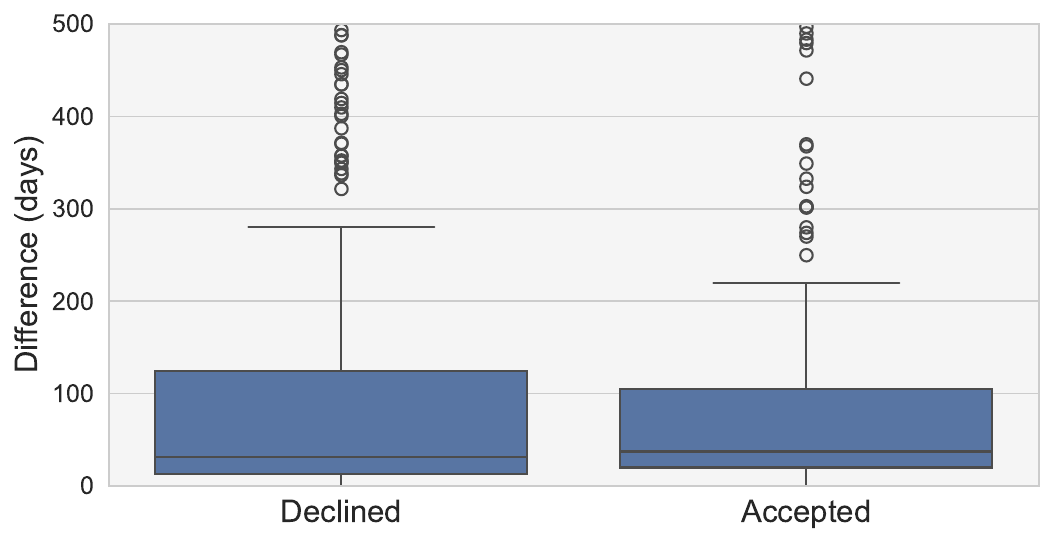}
  \caption{Resolution time of declined and accepted proposals}
  \label{fig:rq2:resolution-time}
\end{figure}

\subsection{Results}
\observation{A larger proportion of proposals are declined (40.9\%) than are accepted (30.8\%).} \label{ob:rq1:declined-proposal-proportion}
\tab{tab:rq1:proposal-statuses} shows the distribution of proposals by state, revealing that declined proposals constitute the largest class (40.9\%), surpassing the acceptance rate (30.8\%).
% Accepted proposals follow at 30.78\%. 
% This finding reveals that most of the proposals are declined, while accepted proposals make up less than one-third of the total.
The high decline rate suggests that the proposal workflow involves careful review and decision-making, which may present challenges for contributors aiming to get their proposals accepted.
Additionally, contributors may violate explicit or implicit criteria, resulting in a higher likelihood of rejection.

\observation{Half of the proposals take more than one month to be accepted or declined.} \label{ob:rq1:resolution-time}
%% ============================================================================
%% PLEASE DON"T REMOVE THIS COMMENT
%% ============================================================================
%% exp20/figures/proposal-resolution-days.pdf
%% exp20/01_01_resolution_time.ipynb
%% ============================================================================
%% ============================================================================	
\fig{fig:rq2:resolution-time} illustrates the distribution of resolution times for both declined and accepted proposals. The median resolution times are 31 days for declined proposals and 37 days for accepted proposals, while the 75th percentile values extend to 124 days and 104 days, respectively. Hence, once a proposal is submitted, it takes a considerable amount of time to reach a decision, regardless of whether it is accepted or declined.
%This indicates that even proposals that are eventually declined often undergo lengthy discussions and evaluations, which suggests that time and effort could be saved if proposals likely to be declined were identified earlier.
%If contributors could anticipate the likelihood of their proposals being declined, they could refine their ideas in advance or avoid submitting proposals unlikely to be accepted. 
%Likewise, maintainers could prioritize reviewing proposals with a higher likelihood of acceptance, making the process more efficient.

\observation{Only 14.7\% of declined proposals have been resubmitted.} \label{ob:rq1:reappeared-proposal}
%\tab{tab:rq1:reappeared-proposal-results} presents the proportion of declined proposals that were referenced in later issues, indicating resubmission. The results show that the minority of declined proposals (14.7\% (66/448)) were resubmitted as new proposals, suggesting that once a proposal is declined, contributors are unlikely to submit a revised version.
%%the majority (over 80\%) of declined proposals were not resubmitted
%This finding implies that, in most cases, proposal declines lead to abandonment rather than further refinement and resubmission.
We compute the proportion of declined proposals that were referenced in subsequent issues, indicating resubmission. The results show that a minority of declined proposals (14.7\% (66/448)) have been resubmitted as new proposals, suggesting that once a proposal is declined, contributors are unlikely to submit a revised version.
This finding implies that, in most cases, a proposal being declined leads to abandonment rather than further refinement and resubmission.

\observation{Over half of the resubmitted proposals are declined.} \label{ob:rq1:reappeared-proposal-results}
\tab{tab:rq1:reappeared-proposal-results} presents the outcomes of resubmitted proposals.
Approximately 53.3\% of resubmitted proposals have been declined again, while only 25.3\% were accepted.
The remaining proposals are still Works In Progress (WIP).\footnote{Rounding down repeating decimals results in a total slightly below 100\%.}
This suggests that resubmission does not always improve the chances of acceptance; however, the fact that some proposals eventually become accepted highlights that refinement and persistence can change decisions.

% Note that ``WIP'' indicates cases where developers have not yet made a decision. 
% \mahmoud{The sum of all the numbers should be 100\%, right? 53.3 + 25.3 + 21.3 = 99.9\%. Maybe just make it up by checking the accurate decimals?}. 

\summarybox{Key Insights of RQ1}
{
% A large proportion of proposals are declined, and this decision-making process takes a considerable amount of time. Anticipating declined proposals and understanding their reasons before submission could save time and effort for both contributors and maintainers, thus making the process more efficient.
Declined proposals are common in the Go community, accounting for over 40\% of all proposals.
Proposal decisions often take over a month, regardless of outcome, reflecting a time-intensive review process.
Although a small portion (14.7\%) of declined proposals are resubmitted, more than half of these are declined again, indicating a need to improve the resubmission process to help contributors avoid repeated rejections.}
% More than 40\% of proposals were declined, indicating a significant number of proposals were declined. Also, are all the declined proposals completely abandoned? Our findings suggest that this is mostly true. Only about 14.7\% of declined proposals reappeared, and over half of these reappeared proposals were also declined. Therefore, we can conclude that a significant number of proposals are indeed discarded.

%%% Local Variables:
%%% mode: latex
%%% TeX-master: "../main"
%%% End:

\section{\RQthree{}}
%\section{Qualitative Analysis}
\label{sec:rq3}
%RQ3: \emph{\RQthree{}}

%\subsection{Motivation}
%\label{sec:rq3:motivation}
Our findings in RQ1 reveal that a substantial portion of declined proposals were resubmitted, yet more than half of these resubmitted proposals were declined again. 
This suggests that contributors may lack a clear understanding of the common reasons for proposal rejection, leading to repeated submissions of proposals that face similar issues.

Understanding why proposals are declined is crucial for improving the proposal process. If contributors can identify the reasons behind past declines--especially for proposals similar to their own--they can refine their submissions before review, increasing their chances of acceptance. 
Similarly, maintainers could use a structured framework to provide clearer feedback, reducing redundant submissions and improving overall proposal quality.
This, in turn, can guide contributors in avoiding common pitfalls and strengthening their proposals, ultimately making the proposal process more transparent and efficient.
Therefore, in this RQ, we aim to construct a taxonomy of decline reasons. 
% A well-defined taxonomy would help maintainers systematically classify declined proposals and provide actionable insights to contributors. This, in turn, can guide contributors in avoiding common pitfalls and strengthening their proposals, ultimately making the proposal process more transparent and efficient.
% Our findings in RQ1 show that a non-negligible portion of declined proposals reappear, and half of these are declined once more.
% Providing an understanding of the reasons behind previously declined proposals, particularly those similar to new proposals, is of a paramount importance to help contributors improve their proposals in the future and avoid common issues that lead to the decline of their proposals. 
% For instance, if contributors understand the reasons why similar proposals were declined, they could revise their own proposals, potentially increasing their acceptance rate.
% To facilitate this, our first step will be to construct a taxonomy of reasons for declined proposals.
% If we provide such a taxonomy, maintainers can classify declined proposals. Contributors can also understand why similar proposals were declined and how to revise their own proposals based on these reasons.

\subsection{Approach}
% To the best of our knowledge, no previous studies have constructed a taxonomy of the reasons for declined proposals. 
To construct the taxonomy of decline reasons, we perform an open coding analysis on comments raised during the review of declined proposals to investigate the decline reasons~\cite{krippendorff2018book}.
This approach has been widely employed in prior studies~\cite{hata2019ICSE,wang2021EMSE,chouchen2021SANER,hata2022EMSE,wang2023EMSE,zanaty2018ESEM,hirao2019FSE}.
%\mahmoud{I would cite this book for the coding analysis instead of previous papers: "K. Krippendorff, Content Analysis: An Introduction to Its Methodology. Sage Publications, 2018."}
% Therefore, we used a manual coding approach to create a taxonomy and classify all declined proposals.
% A \emph{code} represents a concept used to classify the data.
% Therefore, we used an open coding and a closed coding approaches to create a taxonomy and classify all declined proposals.
% Open coding is a method of qualitative analysis that classifies data into new concepts, also known as codes~\cite{hirao2019FSE}.
% This coding process creates a \emph{coding guide} that shows the criteria for classifying data for each code.
% Closed coding is a method used to classify data based on an existing coding guide.
% % closed codingはopen codingで得られたコードなどの既存のコードを使ってデータを分類する手法である．
Below, we detail our coding process.

% 我々はOpen codingを用いてGoのプロポーザルのdeclinedの理由を分類するためのinitial coding guideを構築する．
% 次に，closed codingを用いて，Open codingで得られたコードを使ってデータを分類する．2人のコーダがinitial coding guideをもとにコーディングを行いつつ，コーディングガイドを改良する．
% 以下に詳しいプロセスを記した．

%\masa{maybe I'll add a figure to describe the process}

%\input{figures/rq3/code-freq.tex}
%% ============================================================================
%% PLEASE DON"T REMOVE THIS COMMENT
%% ============================================================================
%% exp15/03_make_table.ipynb
%% exp15/tables/freq.csv
%% ============================================================================
%% ============================================================================	

\begin{table}[t]
\centering
\caption{Frequency of the discovered reasons for declined proposals}
\label{tab:rq3:code-freq}
\begin{tabular}{lrr}
\toprule
\multicolumn{1}{c}{\textbf{Code}}&\multicolumn{1}{c}{\textbf{Count}}&\multicolumn{1}{c}{\textbf{Pro(\%)}}\\
\midrule
Poor feasibility&102&22.8\%\\
Deprecated proposals&81&18.1\%\\
Limited use cases&59&13.2\%\\
Breaking Go's principles&51&11.4\%\\
Existing alternatives&46&10.3\%\\
Lack of knowledge&42&9.4\%\\
Duplication&39&8.7\%\\
No consensus reached&26&5.8\%\\
License problem&2&0.4\%\\
\bottomrule
\end{tabular}
\end{table}

\smallskip
\noindent
\textit{Step 1: Develop an initial coding guide.} To systematically identify the reasons for declined proposals, we apply open coding with a saturation criterion~\cite{miles1994book} to construct an initial coding guide.
%\mahmoud{we can cite this for the saturation criteria: "M. B. Miles and A. M. Huberman, Qualitative Data Analysis: An Expanded Sourcebook. Sage Publications, 1994"}
This guide defines the criteria for classifying declined proposals based on the reasons for rejection.

The saturation criterion serves as a stopping condition for discovering new codes. 
Specifically, we stop coding when no new decline reasons emerge in $N=50$ consecutive randomly sampled proposals, as done in previous studies~\cite{hirao2019FSE,zanaty2018ESEM}. Two coders (the first and second authors) conducted the initial coding, both with extensive programming experience (14 and 7 years, respectively).
%\mahmoud{I think I will have 7 years by the time we submit this paper, say around April : )}

During the coding process, both coders independently analyze the proposal discussions (comments), focusing on the rationale for rejection of maintainers.
They iteratively refine and discuss emerging codes until they reach a consensus.
Saturation was achieved after coding 90 proposals, after which an additional 10 proposals were analyzed for validation, resulting in a total of 100 coded proposals in this phase.

\smallskip
\noindent
\textit{Step 2: Refine the coding guide and assess reliability.} To ensure the reliability of the initial coding guide, we conduct a validation phase with a third coder with two years of Go programming experience. 
The first and third coders independently label 20 randomly sampled declined proposals using the initial guide.

We then measure inter-coder agreement using Cohen's Kappa coefficient~\cite{mchugh2012},
%\mahmoud{we can cite this: "M. L. McHugh, “Interrater reliability: The kappa statistic,” Biochemia medica, vol. 22, no. 3, pp. 276–282, 2012."},
where a score $>=$ 0.81 indicates almost perfect agreement~\cite{viera2005understanding}. 
If the agreement was below this threshold, discrepancies were discussed, and the coding guide was refined accordingly. 
This process was repeated in batches of 20 proposals until the agreement score exceeded 0.81.
After four iterations (coding 80 proposals), we achieve a Cohen's Kappa score of 0.876, confirming strong agreement.

\smallskip
\noindent
\textit{Step 3: Apply the final coding guide.}
We use the coding guide to classify all declined proposals, incorporating the 100 proposals coded in Step 1 (for code discovery) and the 80 proposals coded in Step 2 (for validation and refinement).
The same coders from Step 2 independently label each proposal using the final guide. 
If their classifications were consistent, they were recorded as final. 
For inconsistencies, coders discussed and resolved discrepancies to ensure classification accuracy.

\begin{figure}[t]
  \centering
      \includegraphics[width=\columnwidth]{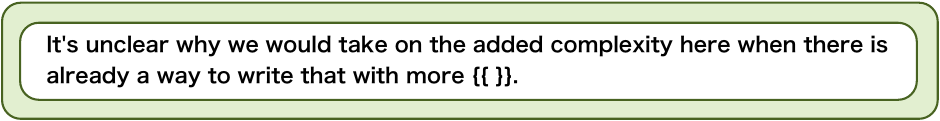}
  \caption{An example declined proposal (R1) retrieved from \texttt{issues/46588} in \texttt{golang/go}~\cite{go-issue-extwo}}
  \label{fig:rq3:example-poor-feasibility}
\end{figure}
%% ============================================================================
%% PLEASE DON"T REMOVE THIS COMMENT
%% ============================================================================
%% ============================================================================
%% ============================================================================	

\begin{figure}[t]
  \centering
      \includegraphics[width=\columnwidth]{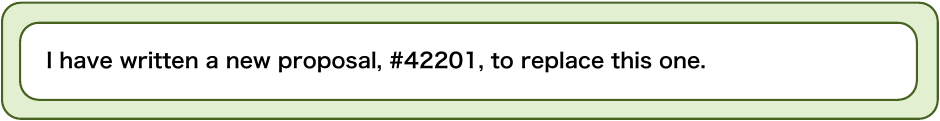}
  \caption{An example declined proposal (R2) retrieved from \texttt{issues/37113} in \texttt{golang/go}~\cite{go-issue-exthree}}
  \label{fig:rq3:example-deprecated-proposals}
\end{figure}
%% ============================================================================
%% PLEASE DON"T REMOVE THIS COMMENT
%% ============================================================================
%% ============================================================================
%% ============================================================================	

\begin{figure}[t]
  \centering
      \includegraphics[width=\columnwidth]{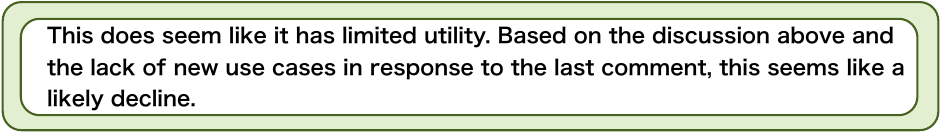}
  \caption{An example declined proposal (R3) retrieved from \texttt{issues/38831} in \texttt{golang/go}~\cite{go-issue-exfour}}
  \label{fig:rq3:example-limited-use-cases}
\end{figure}
%% ============================================================================
%% PLEASE DON"T REMOVE THIS COMMENT
%% ============================================================================
%% ============================================================================
%% ============================================================================	

\begin{figure}[t]
  \centering
      \includegraphics[width=\columnwidth]{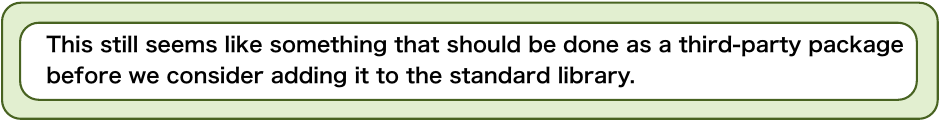}
  \caption{An example declined proposal (R4) retrieved from \texttt{issues/51563
} in \texttt{golang/go}~\cite{go-issue-exfive}}
  \label{fig:rq3:example-breaking}
\end{figure}
%% ============================================================================
%% PLEASE DON"T REMOVE THIS COMMENT
%% ============================================================================
%% ============================================================================
%% ============================================================================	

\begin{figure}[h!]
  \centering
      \includegraphics[width=\columnwidth]{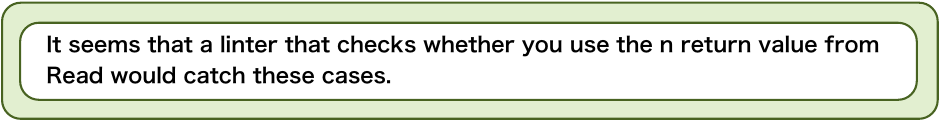}
  \caption{An example declined proposal (R5) retrieved from \texttt{issues/48182} in \texttt{golang/go}~\cite{go-issue-exsix}}
  \label{fig:rq3:example-existing}
\end{figure}
%% ============================================================================
%% PLEASE DON"T REMOVE THIS COMMENT
%% ============================================================================
%% ============================================================================
%% ============================================================================	

\begin{figure}[h!]
  \centering
      \includegraphics[width=\columnwidth]{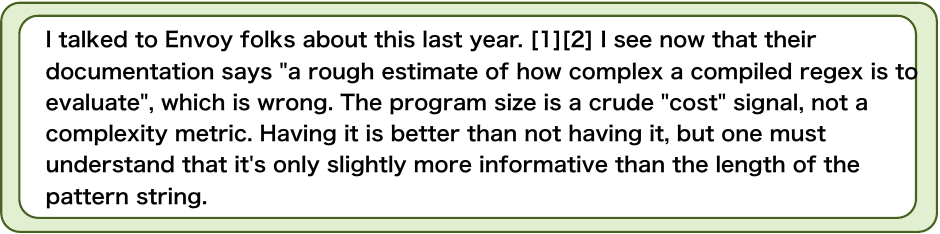}
  \caption{An example declined proposal (R6) retrieved from \texttt{issues/39413} in \texttt{golang/go}~\cite{go-issue-exseven}}
  \label{fig:rq3:example-lack}
\end{figure}
%% ============================================================================
%% PLEASE DON"T REMOVE THIS COMMENT
%% ============================================================================
%% ============================================================================
%% ============================================================================	

\begin{figure}[t]
  \centering
      \includegraphics[width=\columnwidth]{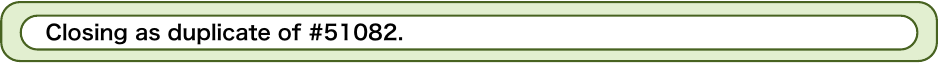}
  \caption{An example declined proposal (R7) retrieved from \texttt{issues/45533} in \texttt{golang/go}~\cite{go-issue-exeight}}
  \label{fig:rq3:example-duplication}
\end{figure}
%% ============================================================================
%% PLEASE DON"T REMOVE THIS COMMENT
%% ============================================================================
%% ============================================================================
%% ============================================================================	

\begin{figure}[t]
  \centering
      \includegraphics[width=\columnwidth]{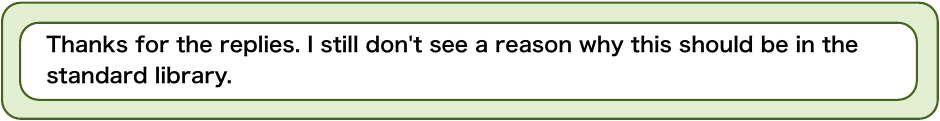}
  \caption{An example declined proposal (R8) retrieved from \texttt{issues/50819} in \texttt{golang/go}~\cite{go-issue-exnine}}
  \label{fig:rq3:example-no-consensus}
\end{figure}
%% ============================================================================
%% PLEASE DON"T REMOVE THIS COMMENT
%% ============================================================================
%% ============================================================================
%% ============================================================================	

\begin{figure}[h!]
  \centering
      \includegraphics[width=\columnwidth]{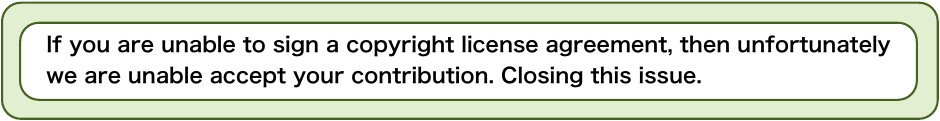}
  \caption{An example declined proposal (R9) retrieved from \texttt{issues/42361} in \texttt{golang/go}~\cite{go-issue-exten}}
  \label{fig:rq3:example-license}
\end{figure}

\subsection{Results}
\observation{Through our analysis, we identified nine main reasons why Go proposals are declined.} \label{ob:rq3:reasons}
\tab{tab:rq3:code-freq} presents an overview of the identified reasons for declined proposals based on our coding analysis.
% The complete table is available in our replication package. 
Below, we describe each reason with examples from proposal discussions.
%\mahmoud{two figures related to Results are placed in the previous page. Can we try to move them down later?}
% Using the process described in \sec{sec:rq3:motivation}, here is the full list of reasons for declined proposals:
\begin{itemize}
  \item
  \textbf{R1. Poor Feasibility (22.8\%):}
  %\mahmoud{can we add the proportion of cases for each reason? (let us add X\% for each reason.}
  % Theoretically possible but generally unfeasible proposals due to high costs or side effects (\eg, ``generally impossible'') are included. Theoretically impossible items are also included here.
  % %Costs or side effects are clearly indicated in the comments. 
  The most common reason for declined proposals is that the proposed change is technically possible but impractical due to high implementation costs, severe side effects, or fundamental limitations. In some cases, the proposal is theoretically impossible. Maintainers typically decline such proposals by citing concerns about the effort required relative to the benefits.
  For example, in \fig{fig:rq3:example-poor-feasibility}, a maintainer explains that while a proposal is feasible, its advantages do not outweigh the costs of implementation.
  %\mahmoud{we can use the label of the reason (i.e., Rx) in the caption of the figure to make them more concise. To get an idea, you can check Fig 3 and 4 in this papae: "On the Use of Dependabot Security Pull Requests"}
  % This category includes proposals that are theoretically possible but generally unfeasible due to high costs or severe side effects (\eg, those labeled as ``generally impossible'' in their comments). It also encompasses theoretically impossible items. Comments should clearly indicate the associated costs and side effects.
  % \fig{fig:rq3:example-poor-feasibility} shows a comment from a maintainer that explains why a proposal was declined.
  \item
  \textbf{R2. Deprecated Proposals (18.1\%):}
  Proposals are declined because they relate to deprecated features or have become obsolete. This includes cases where: 1)  The feature being proposed has already been deprecated; 2) The feature is covered by another active proposal and will be merged; 3) The discussion stalled for an extended period with no further engagement.
  % This is a deprecated proposal. For example, (1) a feature related to this proposal has been deprecated, (2) a feature in this proposal has been proposed in another and will be merged, or (3) a discussion has been abandoned and no further discussion has occurred for a long time.
  \fig{fig:rq3:example-deprecated-proposals} shows an example where a maintainer declines a proposal due to redundancy with a new change. These cases suggest that contributors may lack visibility into ongoing or past discussions.
    %\mahmoud{I am adding some description for each example to bring a context to each reason.}
  % \fig{fig:rq3:example-deprecated-proposals} shows a comment from a maintainer that explains why a proposal was declined.
  \item
  \textbf{R3. Limited Use Cases (13.2\%):}
  Proposals that apply to a very narrow set of use cases are often declined in favor of existing workarounds. Maintainers typically argue that the Go language should prioritize general-purpose solutions over highly specific features that benefit only a small subset of users. \fig{fig:rq3:example-limited-use-cases} shows an example where a maintainer simply suggests the proposed solution lacks use cases. 
  % using an alternative workaround approach instead of implementing the proposed feature.
  This highlights the importance of understanding Go’s design philosophy before proposing changes.
  % The applicable situations and use cases are very limited, so it's better to use another workaround to solve the problem. 
  % \fig{fig:rq3:example-limited-use-cases} shows a comment from a maintainer that explains why a proposal was declined.
  \item
  \textbf{R4. Breaking Go’s Principles (11.4\%):}
    Go maintains a strong design philosophy, including principles such as simplicity, backward compatibility, and a minimal standard library. Proposals that contradict these principles are often declined, even if they introduce useful features.
    For example, in \fig{fig:rq3:example-breaking}, a maintainer rejects a proposal due to its impact on the minimal standard library policy. 
    % its impact on Go’s backward compatibility policy. 
    Contributors unfamiliar with these core principles may unknowingly propose changes that conflict with Go’s long-term goals.
  
  % The proposal breaks Go's philosophy (\eg, design philosophy, standard library promises, and backward compatibility).
  % \fig{fig:rq3:example-breaking} shows a comment from a maintainer that explains why a proposal was declined.
  %\masa{add ex}
  \item
  \textbf{R5. Existing Alternatives (10.3\%):}
  Many proposals are declined because alternative solutions already exist, and contributors are unaware of them. Maintainers often point out that the requested functionality can be achieved using current language features, libraries, or best practices.
    \fig{fig:rq3:example-existing} illustrates a case where a proposal was declined because an equivalent solution was already available. This suggests that better documentation or guidance on existing features could help contributors avoid redundant proposals.
  % Alternative solutions (\eg, ways to solve the problem with the current features) already exist.
  % The contributor is unaware of them. 
  % \fig{fig:rq3:example-existing} shows a comment from a maintainer that explains why a proposal was declined.
  %The proposer was unaware of them, but they were pointed out in the proposal. Therefore, there is no need to accept this proposal.
  \item
  \textbf{R6. Lack of Knowledge (9.4\%):}
  %The proposer has incorrect knowledge of the premise, and the discussion in the proposal is not an argument but a pointing out of the mistake. However, this does not apply to cases where the proposer was unaware of existing alternatives.
  Proposals are based on misconceptions about Go’s behavior, leading to discussions that simply correct misunderstandings rather than evaluating new ideas. However, when a proposal is declined due to an unawareness of existing alternatives, it is classified under the R5 reason instead.
  In \fig{fig:rq3:example-lack}, a maintainer clarifies that the proposed change is unnecessary due to a misunderstanding of a complexity metric.
  This may suggest that educational resources could reduce the number of proposals stemming from incorrect assumptions.
  % The contributor misunderstands the premise, and the discussion in the proposal merely points out this misunderstanding. Note that when the contributor is unaware of alternative solutions, the proposal is classified under existing alternatives. 
  % \fig{fig:rq3:example-lack} shows a comment from a maintainer that explains why a proposal was declined.
  
  \item
  \textbf{R7. Duplication (8.7\%):}
    Duplicate proposals--those that overlap with previously submitted ideas--are often declined. In some cases, maintainers merge discussions, but in others, they simply close the redundant proposal. For instance, in \fig{fig:rq3:example-duplication}, a maintainer points out that a similar proposal already exists. These cases highlight the importance of searching for related proposals before submission.
  % Other proposals exist that contain the same content. 
  % \fig{fig:rq3:example-duplication} shows a comment from a maintainer that explains why a proposal was declined.
  \item
  \textbf{R8. No Consensus Reached (5.8\%):}
  %This proposal could be a subject of discussion. However, it was not adopted because
  Proposals that lack a clear agreement among maintainers, contributors, and the community are typically declined. This happens when discussions remain inconclusive due to divergent opinions on necessity or implementation details. 
  \fig{fig:rq3:example-no-consensus} shows an example where a proposal was declined because maintainers could not reach an agreement. This suggests that stronger justifications and well-supported arguments can improve a proposal’s chances of acceptance.
  % There was no clear consensus on its necessity or implementation among all participants.
  % \fig{fig:rq3:example-no-consensus} shows a comment from a maintainer that explains why a proposal was declined.
  %, including the contributor.
  \item
  \textbf{R9. License problem (0.4\%):}
  Proposals are also declined due to licensing issues, such as contributors not consenting to the required copyright terms. 
  In \fig{fig:rq3:example-license}, a maintainer closes a proposal for this reason. These cases are less common but highlight the importance of legal compliance in open-source contributions.
  % The contributor did not give consent for the contribution, such as accepting copyright.
  % \fig{fig:rq3:example-license} shows a comment from a maintainer that explains why a proposal was declined.

\end{itemize}
 
%% ============================================================================
%% PLEASE DON"T REMOVE THIS COMMENT
%% ============================================================================
%% exp15/02_for_sheet.ipynb 
%% exp15/csv/output.csv
%% ============================================================================
%% ============================================================================	

%Of 100 declined proposals, 8 were declined for two reasons. Thus, the total frequency count is 108.
%The full table is available online.\footnote{\url{https://docs.google.com/spreadsheets/d/1ra01NCAOKTE4oqW0xsimzxiImCu6DUk__stgQBd-5T0/edit?usp=sharing}}

Overall, the most frequent reason for declining proposals is poor feasibility, which often demands a deep technical understanding of trade-offs, implementation complexity, and system constraints. Since maintainers have broader experience with Go’s long-term direction, contributors may find it difficult to anticipate these concerns in advance.
Similarly, other reasons, such as breaking Go’s principles, limited use cases, and existing alternatives, require contributors to be well-versed in Go’s design philosophy and ecosystem. This knowledge gap between contributors and maintainers highlights the importance of tool support to help contributors assess their proposals before submission.
%\mahmoud{Thanks for your text below. I used it to re-frame the overall observation of this RQ, but I feel we don't really need a very specific observation here. The observation can be finding these reasons themselves. What do you think? If you don't like it, feel free to uncomment what I commented please : )}
% \observation{Understanding the reasons for declined proposals is often challenging for contributors because it requires in-depth knowledge.}
% The most common reason for declined proposals is ``Poor feasibility.''
% For example, \fig{fig:rq3:example-poor-feasibility} shows a declined proposal due to poor feasibility. The maintainer commented that the proposal was possible but its benefits were not worth the costs.
% Evaluating this type of proposals before submitting them to maintainers requires contributors to fully understand their advantages and disadvantages. However, typically only maintainers possess such comprehensive knowledge. 
%Decisions may also depend on implementation priorities, which largely depend on the judgment by maintainers.

% Other categories similarly demand an in-depth understanding of Go. The need for contributors to have an in-depth understanding of Go as much as maintainers highlights the importance of developing tool support to bridge the knowledge gap between contributors and maintainers.

\summarybox{Key Insights of RQ2}
{
% There are various reasons for declining proposals, most of which require in-depth knowledge of Go. Therefore, we need to support contributors in understanding why their proposals might be declined.
% Therefore, providing better support mechanisms can help contributors understand potential rejection reasons early, improving proposal quality and alignment with Go’s development priorities.
The reasons for declining proposals range from technical feasibility challenges (R1, R3, R5) and violations of Go’s design principles (R4) to community-driven factors such as lack of consensus (R8) and licensing issues (R9).
%\mahmoud{part of this summary can be in the implication discussion (the tool support. Here, let us maybe focus on the findings only. I rephrased it a bit to do that. In the implication, we can discuss that this result shows that providing better support mechanisms can help contributors understand potential rejection reasons early, improving proposal quality and alignment with Go’s development priorities.}
}

\section{\RQtwo{}}
%\section{Inferring Declined Proposals}
\label{sec:rq2}
%RQ2: \emph{\RQtwo{}}

%\input{figures/rq2/rq2_overview.tex}
%\input{tables/rq2/likely-to-final.tex}

%\masa{2024/10/31, exp17: request-1458, num comments: 5, issue: 48152 results in pending review. So we consider this prediction is false (declined). we need to describe this. please check the first cell in 02\_07 ipynb}\masa{Fixed by 2024/11/1}

%\subsection{Motivation}
% declined proposalsをdecision makingの前に判定することができればメリットがある．メンテナは判断をする手間を必要としなくなる．貢献車も判定が行われるまでの時間を無駄にしなくて済む．このRQでは，declined proposalsを判定するためのモデルを構築する．
% 図1に問題設定を図解した．Section 3で説明した通り，Goにおけるproposalsの判定はいくつかのフェーズに分かれている．そのうち，likely declined/acceptedの判定が下されるところまでに，そのproposalsがdeclinedになるかacceptedになるかを予測するモデルを作成する．これはlikely declinedやlikely acceptedになったproposalsの判定が覆ることはほとんどないためである（Table X）．そのため，likely acceptedなどの判断を下すところのデータも使った場合，その判定と同じ判定を最終判断とするだけで予測モデルの性能が良くなってしまう．
% 図1に示すようにlikely accepted/declinedの判定が下されるところまでに行われるProposalの説明およびコメントを入力とする．GPT-3.5-turboにこれらのデータを入力し，最終的な判断を予測させる．Proposalsの総数はXXX件であり，教師あり学習を学習させるには不十分なデータ数であるため，pre-trained済みのモデルを使用する．
% 我々の仮定は，Proposalsの判定が下される直前までのコメントを利用した場合に予測精度が高くなり，Proopsalが投稿されてからできるだけ早いタイミング，つまりproposalのdescriptionのみを使いコメントを一切使わない場合に，その予測の確率が低くなるというものである．一方で，判断が下される直前での予測は開発者および貢献者の議論が十分行われた後であり，開発者をサポートするという意味では最小限の貢献である．コメントが0件でProposalの説明文のみの段階で予測できた場合に，これは最も有用なモデルである．
% そこで，図2に示すように，各プロポーザルにおけるコメントの割合ごとに予測精度を比較する．コメント数はproposalsごとに異なるため，各proposalにおいてコメントの割合を算出する．

The Go proposal process requires maintainers to review and discuss each submission before making a decision. 
In fact, RQ1 shows that half of the proposals take more than one month to reach a decision. Moreover, 25\% of the proposals take more than three months. This long review time can be a burden for both contributors and maintainers.
Contributors invest effort in crafting proposals, often without guarantee that their proposals will be accepted. In fact, a large proportion of proposals are declined (40.9\%).
Similarly, maintainers must dedicate considerable time to evaluating proposals, even those that are ultimately declined.
%Given that more than 40\% of the proposals are declined, this process can be time consuming and resource intensive for both contributors and maintainers.
%Contributors invest effort in crafting proposals, often without clear guidance on whether their ideas align with Go’s development priorities.

Even proposals that are eventually declined often undergo lengthy discussions and evaluations, which suggests that time and effort could be saved if proposals likely to be declined were identified earlier.
If contributors could anticipate the likelihood of their proposals being declined, they could refine their ideas in advance or avoid submitting proposals that are unlikely to be accepted. 
Likewise, maintainers could prioritize reviewing proposals with a higher likelihood of acceptance, making the process more efficient.
Therefore, in this RQ, we explore whether declined proposals can be predicted at an early stage.
%RQ4 will explore whether declined reasons can be predicted.  
%and declined reasons

% \mahmoud{Please check this version of motivation if you like it!}
% \masa{Thank you! I really like this version.}
% It would be beneficial if contributors could independently identify declined proposals without requiring decisions from maintainers. This approach would also reduce the workload of maintainers by avoiding the need for proposal discussions. Also, contributors could identify potential improvements in their proposals at an earlier stage.
% To achieve this, this RQ aims to develop a prediction model for identifying potentially declined proposalsa at an early stage.

% \fig{fig:rq2:resolution-time} illustrates the distribution of resolution times for both declined and accepted proposals. The median resolution times for declined and accepted proposals are approximately 31 days and 37 days, respectively. Similarly, the 75th percentile resolution times are around 124 days and 104 days. These findings suggest that a prediction model could significantly support both maintainers and contributors by reducing the time spent on proposals that are likely to be declined.

%% ============================================================================
%% PLEASE DON"T REMOVE THIS COMMENT
%% ============================================================================
%%
%% ============================================================================
%% ============================================================================	

\begin{figure}[t]
  \centering
      \includegraphics[width=\columnwidth]{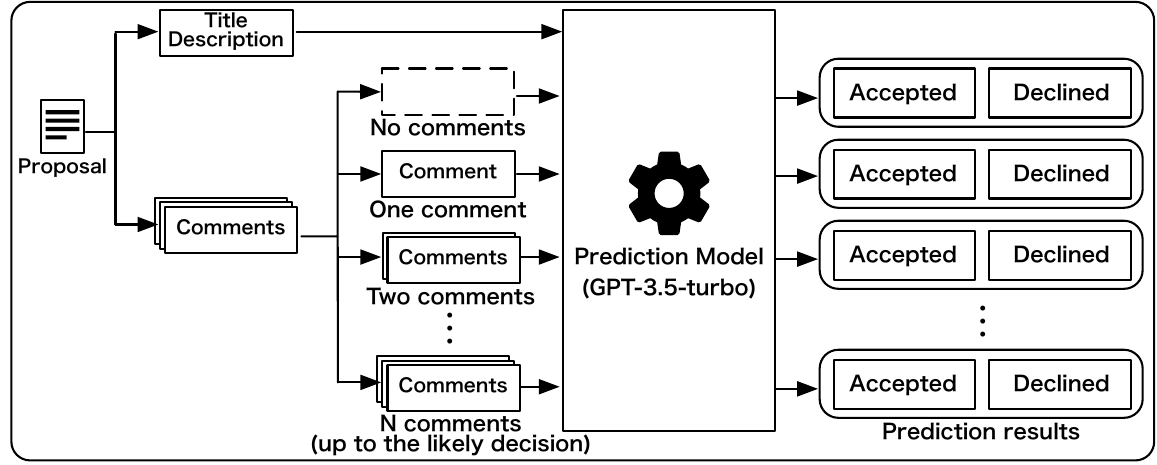}
  \caption{Prediction model for determining accepted and declined proposals}
  \label{fig:rq2:prediction}
\end{figure}

\subsection{Approach}
\fig{fig:rq2:prediction} illustrates our approach for predicting whether a proposal will be declined or accepted before the ``likely declined/accepted'' decision is made.
% Note that final decisions are rarely reversed once a proposal reaches this stage, and hence, early prediction can help contributors refine proposals before extensive discussions take place.
We build a prediction model using GPT-3.5-turbo, which takes proposal descriptions and discussion comments (only those available at the time of the prediction) as input, and predicts the outcome of the proposal.
Our input prompt can be found in our replication package.
% \fig{fig:rq2:prediction} illustrates our prediction model for identifying declined proposals. As explained in \sec{sec:back:proposal}, the decision process on proposals in Go involves multiple phases. We aim to construct a model to predict whether a proposal will be ultimately declined or accepted before the ``likely declined/accepted'' decision is made. This approach is justified by the rarity of final decisions being reversed once a proposal is judged as likely declined or accepted.
%% ============================================================================
%% PLEASE DON"T REMOVE THIS COMMENT
%% ============================================================================
%% exp17/01_01_create_db.ipynb
%% ============================================================================
%% ============================================================================	
% Indeed, among 337 accepted proposals, 3 cases included keywords associated with a declined decision (\eg, likely declin), while among 448 declined proposals, 44 cases included keywords indicating an accepted decision (\eg, likely accept).
% To alleviate the burden on contributors and maintainers, it is crucial to employ the model for predictions prior to the ``likely declined/accepted'' decision.
% \mahmoud{I don't think we need this info, so I commented it.}

The effectiveness of the prediction model depends on the input data. Using all comments up to the likely accept or decline decision is impractical, as discussions would have already shaped the outcome. In contrast, using only the title and description makes the model suitable for early-stage guidance. 
To examine the impact of different input sizes, we test the model with varying amounts of comment data from only the title and description to all comments before the likely decision.
% (\fig{fig:rq2:example-acc}).

%\input{figures/rq2/example-acc}

For the model input, we initially select the 735 proposals that received a final decision (334 accepted, 401 declined). To ensure the model only relies on information available at the time of prediction, we include only proposals where the likely accept/decline decision was made after the last training date for GPT-3.5-turbo (September 30th, 2021). This ensures that the model would not have encountered the studied discussions or proposal outcomes during its training process.
After applying this filter, the final dataset consists of 260 proposals (118 accepted, 142 declined).

% \smallskip
% \noindent
% \textbf{Step 1: Identifying Decision Points.}
% To determine proposal status, we use decision-related keywords (likely accept, likely decline, accept, decline), as detailed in Section~\ref{sec:design}. 
% We consider the earliest occurrence of these keywords as the decision point and used all comments before this point as input for prediction. This process yielded 735 proposals (334 accepted, 401 declined).

% \smallskip
% \noindent
% \textbf{Step 2: Filtering by Model Training Period.}
% To ensure the model does not rely on data beyond its training period, we excluded proposals where the decision was made before October 1, 2021 (GPT-3.5-turbo’s training cutoff was September 2021). This resulted in a final dataset of 260 proposals (118 accepted, 142 declined).

We evaluate the performance of the model using:

\begin{itemize} 
%\item \textbf{Accuracy (Accepted)}: The proportion of correctly identified accepted proposals out of all actual accepted proposals.
    %Measures how well the model identifies accepted proposals.\mahmoud{Can we elaborate more on this, how it is computed? Similar to the text in the metrics below.}
\item \textbf{Precision}: The proportion of correctly predicted declined proposals out of all predicted declines.
\item \textbf{Recall}: The proportion of correctly identified declined proposals out of all actual declined proposals.
\item \textbf{F1 Score}: The harmonic mean of precision and recall for declined proposals.
\end{itemize}

%Since we treat declined proposals as the positive class, Recall effectively measures how well the model identifies them. Therefore, we report Recall instead of Accuracy (Declined), as they are equivalent in this context.

%\input{figures/rq2/evaluation.tex}
%% ============================================================================
%% PLEASE DON"T REMOVE THIS COMMENT
%% ============================================================================
%% exp17/06_01_evaluation.ipynb
%% exp17/tables/evaluation-percentiles.csv
%% OLD versions
%% exp17/05_01_makeplot.ipynb
%% exp17/tables/evaluation.csv
%% OLD versions
%% exp17/05_01_makeplot.ipynb
%% exp17/plot/evaluation.pdf
%% ============================================================================
%% ============================================================================	

\begin{table}[t]
\centering
\caption{Prediction performance based on all comments (All), no comments (NoComments), 25th, 50th, and 75th percentiles prior to the decision}
\label{tab:rq2:evaluation}
\begin{tabular}{lrrrrrr}
\toprule
\multicolumn{1}{c}{\textbf{Metrics}}&\multicolumn{1}{c}{\textbf{NoComments}}&\multicolumn{1}{c}{\textbf{25p}}&\multicolumn{1}{c}{\textbf{50p}}&\multicolumn{1}{c}{\textbf{75p}}&\multicolumn{1}{c}{\textbf{All}}&\multicolumn{1}{c}{\textbf{Diff}}\\
\midrule
F1&0.669&0.715&0.711&0.752&0.875&0.206\\
Precision&0.718&0.698&0.679&0.718&0.821&0.103\\
Recall&0.627&0.732&0.746&0.789&0.937&0.310\\
\bottomrule
\end{tabular}
\end{table}

\subsection{Results}
% \mahmoud{For the results, I am thinking about the following questions/attacks that reviewers might ask when reading the results of RQ2:
% - Why the model performs well for some part of the data (proposals) and not the others. Is there some explanation we can use to explain for example why there are still a few cases the model couldn't anticipate as declined. Also, why the model does not work well for accepted proposals? Can we do some manual analysis on a sample of each case and try to find some reasons? Maybe there are factors like the size of the discussions in correctly predicted cases vs. not correctly predicted cases.}
% \mahmoud{How many cases like that? Can we count? Reviewers may also ask a question: why this is the case?}
% \masa{I have added an additional implication (Implication 6). What do you think?}
% プロポーザルにおける議論はdeclined proposalsを発見するのに十分な情報を提供する．
% 図4にはlikelyの判断が下される前の全てのコメントを利用した時の予測結果（All）およびコメントを使わずにProposalのタイトルと説明文のみを使った時の予測結果（NoComments）を示した．Allは全ての評価指標で0.8以上の値を示している．特に，Recallは1.00となっている．positiveデータはdeclined proposalsであるため，declined proposalsを見逃すことなく捉えることができている．Table 4のConfusion matrixを見ると，実際にdeclined proposalsをAcceptedと予測するケースが0件で会ったことがわかる．この結果は，予測モデルがdeclined proposalsをほとんど見逃さないことを示している．
\observation{Proposal discussions provide crucial information for predicting declined proposals.} \label{ob:rq2:gap}
\tab{tab:rq2:evaluation} shows the prediction performance of the model under five conditions: 
\begin{itemize}
    \item \textbf{All}: Using all comments before the likely accepted/declined decision. This shows a theoretical upper limit on model performance, but has limited practical value. 
    \item \textbf{NoComments}: Using only the title and description without comments. This is the most practical setting, but theoretically shows the lowest performance of the model.
    \item \textbf{25p, 50p, 75p}: Using comments up to the 25th, 50th, and 75th percentiles of the comments before the likely decision. These show how quickly the model acquires enough data to make an accurate prediction without waiting until all comments have appeared. 
\end{itemize}
% The difference in prediction performance across these conditions is significant. Particularly, the gap between the All and NoComments conditions is substantial, 0.206 in F1 score, 0.103 in Precision, and 0.310 in Recall.
% Hence, the discussion provides crucial information for anticipating declined proposals.
The differences in prediction performance across these conditions are substantial. In particular, the gap between the All and NoComments conditions amounts to 0.206 in F1 score, 0.103 in precision, and 0.310 in recall.
When using all available comments, the model achieves high performance across all metrics (more than 0.82).
In particular, the recall for declined proposals reaches 0.93, indicating that the model correctly identifies nearly all declined proposals. 
In contrast, the NoComments condition in \tab{tab:rq2:evaluation} shows a drop in performance across all metrics (less than 0.72). 
These results demonstrate that discussion content is critical for anticipating declined proposals.

\begin{figure}[t]
  \centering
      \includegraphics[width=\columnwidth]{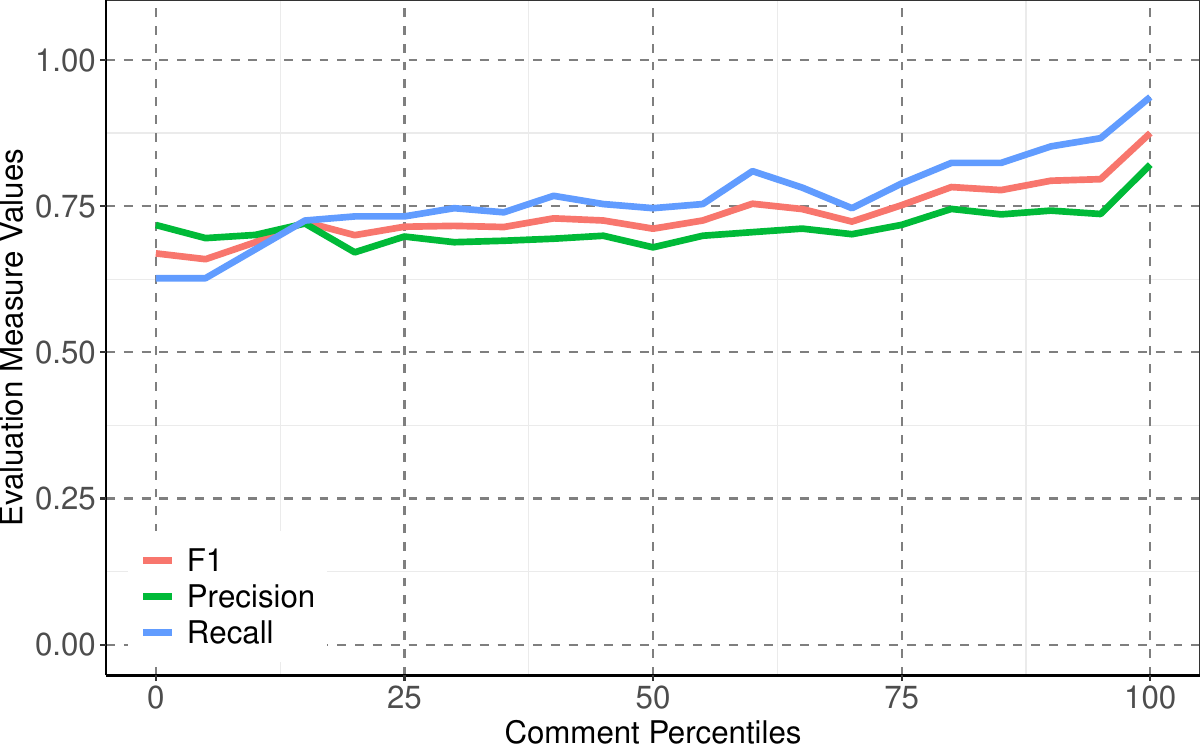}
  \caption{Differences in performance across various percentiles of comments}
  \label{fig:diff-across-percentiles}
\end{figure}

% 議論が行なわれていない段階での予測精度は低いが，議論が進むと予測精度が向上する傾向にある．
% 図4のNoCommentsを見ると全ての評価指標で0.7未満となっている．そのため，proposalのtitleとdescriptionのみではGPTによる予測がうまくいかない可能性がある．Table 5はNoCommentsに対するconfusion matrixである．この表からもaccepted/declined proposalsの双方に対して多くのFalse positiveおよびFalse negativeが発生していることがわかる．ただし，コメントの割合が増えるにつれて予測精度が向上している．図5はx軸にそのProposalsでlikely accepted/declinedのdecisionsまでに行われたコメントの割合を示し，y軸に予測精度を示している．全ての指標で概ね右肩上がりになっている．ただし，コメント数は増えるが予測精度が下がる場合もある．
\observation{The model struggles to distinguish between accepted and declined proposals when using only the title and description but tends to improve as discussions progress. However, their improvement is not always consistent.} \label{ob:rq2:improvement-inconsistency}
As discussions evolve, prediction accuracy improves (\fig{fig:diff-across-percentiles}).
%The 25th percentile condition (25p) and 50th percentile condition (50p) achieve F1 scores of 0.71, respectively; the 75th percentile condition results in a F1 score of 0.75.
The 25th and 50th percentile conditions each achieve an F1 score of 0.71, while the 75th percentile condition reaches 0.75.
The 25th percentile improves the model performance compared to the NoComments condition (0.66).
However, the performance improvement is not changed much across the 25th and 75th percentiles. 
The 75th percentile provides further improvement, although a gap remains with the All condition (0.75 versus 0.87 in F1 score).

% but its improvement is limited until the 50th percentile condition (i.e., the 25th and 50th percentiles show similar performance, 0.71 in F1 score). The 75th percentile condition shows a further improvement, but the gap between the 75th and All conditions is not negligible (0.75 vs. 0.87 in F1 score).
% This indicates that adding the first 25th percentile of comments yields a substantial improvement over the NoComments condition, but gains plateau through the 50th percentile.

% Based on these results, we conclude that the prediction without comments (NoComments condition) should be avoided, as it leads to a suboptimal performance. When incorporating comments, leveraging only the first 25th percentile of comments provides a reasonable trade-off between cost and benefit, since extending to the 50th or 75th percentile delivers negligible further gains. Unsurprisingly, using all comments (All condition) yields the most accurate predictions, but it is impractical for early-stage predictions, as it requires waiting for all discussions to conclude.
% Note that we discuss the cases where an increase in comments does not always lead to improved prediction accuracy in \sec{sec:discussion:researchers}. 

Based on these results, we conclude that the NoComments condition should be avoided, as it leads to suboptimal performance. Incorporating only the first 25th percentile of comments offers a reasonable trade-off between cost and benefit, as extending to the 50th or 75th percentile produces negligible additional gain. Unsurprisingly, using all comments (All condition) yields the highest accuracy, but it is impractical for early predictions because it requires waiting for the entire discussion.
Note that we discuss the cases where an increase in comments does not always lead to improved prediction accuracy in \sec{sec:discussion:researchers}. 

% \fig{fig:diff-across-percentiles} illustrates that performance increases as more comments are added. 
% This suggests that early-stage proposal descriptions alone often lack sufficient information for accurate classification, while discussions provide additional context that helps differentiate between accepted and declined proposals.
% Note that there are cases where an increase in comments does not always lead to improved prediction accuracy.
% For example, the 50th percentile condition shows a slight decrease in performance compared to the 25th percentile condition in terms of precision and F1 score.}
% We discuss these cases in \sec{sec:discussion:researchers}. 

% This variability could be attributed to factors such as redundant discussions, conflicting opinions, or cases where the proposal's core idea remains unclear despite additional comments.
% The NoComments in \fig{fig:rq2:evaluation} shows that the prediction performance falls below 0.72 for all evaluation criteria. \tab{tab:rq2:confusion_nocomments}, the confusion matrix for NoComments, reveals numerous false positives and negatives for both accepted and declined proposals. However, prediction accuracy improves as the percentage of comments increases. \fig{fig:diff-across-percentiles} illustrates this trend, with the x-axis showing the percentage of comments made for each proposal until a likely accepted/declined decision, and the y-axis showing prediction performance. Note that there are cases where an increase in comment numbers does not correspond to improved prediction accuracy.

\summarybox{Key Insights of RQ3}
{
% Proposal discussions substantially improve prediction performance, with the model achieving 0.93 recall for declined proposals when using all comments\textcolor{red}{, whereas it drops below 0.72 when using only the title and description.
% Declined proposals can be anticipated but it requires rationale from not only the proposal title and description provided by the contributor but also the discussions among maintainers and contributors.
% The prediction performance gap between using all comments and only the title and description is substantial (0.206 in F1 score, 0.103 in Precision, and 0.310 in Recall), indicating that the model struggles to distinguish between accepted and declined proposals when using only the title and description.
% As discussions evolve, prediction accuracy improves but it is not always consistent. Our finddings indicate that leveraging only the first 25th percentile of comments provides a resonable trade-off between cost and benefit, as extending to the 50th or 75th percentile yields negligible further gains.
The performance gap between using all comments and using only the title and description is substantial (0.206 in F1 score, 0.103 in precision, and 0.310 in recall), indicating that the model struggles to distinguish between accepted and declined proposals when only the title and description are available.
As discussions evolve, prediction accuracy generally improves, although not always consistently. Our findings indicate that leveraging only the first 25th percentile of comments offers a reasonable trade-off between cost and benefit, since extending to the 50th or 75th percentiles yields negligible additional gains.
% Using all discussion comments, the model predicts declined proposals with high accuracy (recall = 0.93, precision = 0.82), rarely missing actual declined cases.
% In contrast, using only the title and description reduces performance ($<$ 0.72 across metrics), with many false positives and negatives. 
% Prediction accuracy improves as discussions progress, suggesting that proposal discussions contain key signals for anticipating decline.
}

%Without comments (NoComments condition), performance drops below 0.72, increasing false positives and false negatives.
% GPT-3.5-turbo can anticipate declined proposals. However, even for GPT, it is challenging to predict declined proposals without using the discussions within the proposal. 

%\input{sections/rq4}
\section{Discussion}
\label{sec:discussion}
In this section, we discuss the implications of conducting our empirical study.
%\subsection{Implications}
%\masa{I guess we need to discusss the diff between PRs (code-level decisions) and proposals (design-level decisions).}
\subsection{Implications for Maintainers}
% \implication{Using a prediction model at an early stage of proposal discussion can reduce workload on maintainers, but the model performance still needs to be improved.}
\implication{Our model can assist with early-stage triage by identifying proposals that are likely to be declined based on initial discussions.}
Our results show that the model achieves an F1 score of 0.71 using only the first quarter of discussion comments, indicating its potential to anticipate proposal outcomes early in the discussion. 
This early-stage prediction capability can support triage by helping maintainers prioritize proposals likely to be accepted and reduce time spent on those likely to be declined. 
Given that only 30.8\% of proposals are ultimately accepted (Observation~\ref{ob:rq1:declined-proposal-proportion}) and that over half of resubmitted declined proposals are rejected again (Observations~\ref{ob:rq1:reappeared-proposal} and~\ref{ob:rq1:reappeared-proposal-results}), such triage can facilitate review efforts and improve efficiency.\\

\subsection{Implications for Contributors}
% \implication{Predicting whether a proposal will be declined and providing reasons for the decline at an early stage of discussion would help contributors improve their proposals and prevent them from being discouraged.}
\implication{Our prediction models can help contributors revise proposals by identifying likely decline outcomes.}
Contributors often wait weeks for feedback---Observation~\ref{ob:rq1:resolution-time} shows that half of the proposals take over a month to resolve.
This delay may discourage further engagement, e.g., only 14.7\% of declined proposals are ever resubmitted (Observation~\ref{ob:rq1:reappeared-proposal}). 
Hence, providing early predictions could help contributors use that waiting time productively to improve proposals, potentially increasing acceptance rates.
Contributors can also use our declined reasons to understand common patterns in proposal declines. 

\subsection{Implications for Researchers}
\label{sec:discussion:researchers}
%===
% \implication{Predicting reasons for decline requires providing the model with additional context to support complex inference.}
% Observation~\ref{ob:rq4:limit-gpt4o} indicates that even with GPT-4o, the prediction performance for ``Breaking Go's principles'' and ``Lack of knowledge'' remains low.
% Observation~\ref{ob:rq4:limit-gpt35} categorizes these reasons under the complex inference, which requires comprehensive knowledge of the Go project to accurately infer the reasons. This suggests that the prediction model must be provided with additional context information to effectively predict decline reasons. One possible future research direction is to build a database containing the necessary information about the Go project and to use retrieval augmented generation (RAG)~\cite{gao2023arxiv} to predict the reasons for a decline. RAG can access the database and retrieve relevant context to improve prediction accuracy.\\
\implication{Proposing a decline reason prediction model requires enriching model inputs with project-specific context.}
%\implication{Improving decline reason prediction requires enriching model inputs with project-specific context.}
Categorizing declined reasons are more complex than simply predicting whether a proposal will be accepted or declined since (1) it is multiple class classification, and (2) it requires complex inference. For example, the reason ``Breaking Go's principles'' requires a deep understanding of the philosophy and design decisions of the Go project. 
%Observation~\ref{ob:rq4:limit-gpt4o} shows that even with GPT-4o, prediction accuracy for reasons such as Breaking Go's principles and Lack of knowledge remains low. These categories, identified in Observation~\ref{ob:rq4:limit-gpt35} as requiring complex inference, demand a deep understanding of the Go project’s philosophy and design decisions.
This suggests that models need additional context to make accurate predictions. A promising research direction is to apply retrieval-augmented generation (RAG)~\cite{gao2023arxiv}, enabling models to retrieve relevant project knowledge (e.g., Go FAQs, design docs) and improve their reasoning capabilities.
\\
% Declineになった理由を予測する手法を提案するためには，より多くの情報を提供する必要がある

% For example, the discussion between maintainers and contributors would be useful to predict the reasons for the decline. However, such information needs a significant effort of both maintainers and contributors.

% \input{tables/rq2/transition_patterns.tex}
% \input{figures/rq2/change_type.tex}

% \implication{Improving the consistency of prediction results is necessary to improve the prediction accuracy for declined proposals.}
% \mahmoud{I feel this implication has new results, so it is a mix of discussion and implication. In fact, I believe this section should be only about discussing existing results and waht they imply. I'd suggest: 1) Move results to the Results section (Amaybe we add a new subsection: “Prediction Stability Over Time” (if not already there), 2) this implication can be rewritten as below. The transition pattern analysis and the insights about prediction inconsistency are rich and detailed, but they don't quite distill into a standalone actionable implication at the same level as the others.}

\implication{Prediction models must provide stable outputs as proposal discussions evolve.}
% \mahmoud{RQ2 reveals that prediction performance does not consistently improve as more comments are added.
% While some proposals are correctly predicted from the start, many show fluctuating predictions throughout the discussion. Addressing this inconsistency is essential for making models reliable tools in real-time triage and review support.}
Observation~\ref{ob:rq2:improvement-inconsistency} shows that increasing the number of comments does not consistently improve the prediction accuracy for declined proposals. However, Observation~\ref{ob:rq2:gap} indicates a substantial performance gap between using only the title and description and using all comments.
These observations imply that discussion comments are crucial for improving the prediction performance, but the model does not reliably leverage additional comments.
If we develop a method that consistently improves prediction accuracy as the number of comments increases, discussion progress can offer more actionable guidance to maintainers and contributors.

\section{Related Work}
\label{sec:rw}
In this section, we discuss the related work and reflect on how the work compares with ours.\\
% Finally, we review related studies to contextualize our research study (Section~\ref{sec:back:related}).

\noindent \textbf{Studies on Factors Influencing PR Outcomes.}
%\mahmoud{The first two paras are a bit long and I think the message of both is simple: discuss Pull Requests, right? I tried to make it concise below. Also, can we split the section into multiple sections? For example, one clear section could be about "Studies on Factors Influencing PRs Outcome".}
PRs are a GitHub feature for mainly submitting code-level modification requests to OSS projects. Each PR includes a source code patch and description, allowing maintainers to evaluate changes~\cite{pull-requests}. Given the high volume of PRs on GitHub, deciding which to accept/reject is a challenging task~\cite{gousios2015ICSE}.
Therefore, prior studies have focused on understanding the factors influencing PR decisions to improve the development process and support contributors and maintainers. 

There exist several studies that provide insights into key determinants of rejection of PRs~\cite{li2022TSE,gousios2014ICSE,steinmacher2018ICSE}.
Li~\et~\cite{li2022TSE} conducted an empirical study to understand the reason for the abandonment of PRs.
They manually analyzed 321 abandoned PRs from five popular OSS projects on GitHub.
Additionally, they conducted a survey of contributors and integrators, who can review PRs, receiving 619 and 91 responses, respectively. Their analysis identified 12 reasons for the abandonment of PRs.
For example, the most common reason was the lack of answer from integrators. 
%\mahmoud{Can we add the most common reasons or at least briefly list a few of them here?}. 
%However, this study focuses on pull requests abandoned.
%\mahmoud{do you mean declined "proposals"? If so, then it is not needed to mention it since we have one para at the end for this purpose}
% \mahmoud{what do integrators mean here?}
% GitHub上で5つの人気のOSSから放棄された321のプルリクエスト（the review discussion）を集めてきて手作業で分析した．さらにコントリビュータとインテグレータに対してサーベイを実施して，619と91の回答をえた．分析からabondantの12の理由を明らかにした．これは3つのカテゴリに分かれる（pull request，personal，pull requests）．しかし，これは社会的なものであり技術的な理由が少ない．本研究では技術的な理由に焦点を当てて議論している．　
% ただし，この研究は著者によって放棄されたPRについての分析であり，declineされたPRについての分析ではない．introでハイライトした部分に書かれている．Figure 1が模式図だが，これを見てもdeclineの話とabandandの話がちょっと異なることがわかる．
% DONE 読んだ
% 「アクセプトされるprやissueに関する分析」および「見捨てられたPRやissueに関する研究」のディレクトリを参照
Gousious~\et~\cite{gousios2014ICSE} analyzed 350 PRs to investigate the reasons they were closed without being merged. 
%They found that the primary reasons were related to the nature of distributed development rather than technical issues.
They found that the main challenges of closed PRs stemmed from the complexities of distributed development, such as communication barriers, rather than technical issues like bugs or code errors.
%\mahmoud{not clear to me what does "distributed development" means. But maybe we can write it this way if you agree: "They found that the main challenges of closed pull requests stemmed from the complexities of distributed development, such as communication barriers, time zone differences, and collaboration difficulties, rather than technical issues like bugs or code errors.}
% この論文では350のpull requestsを分析して，マージされずに閉じられたpull requestsの理由について分析している．マージされない理由としてtechnical issueよりも分散開発に起因する理由が主なことを述べている．
% なんかこの論文では，マージされる決定に影響を与える要因があることが書かれているらしい．RQ3とRQ4が結構気になるところ．Section 7でmergeの決定に影響を与える要因が分析されている．基本はランダムフォレストモデルでマージされるか否かを予測し，そのモデルでfeature importanceを計算する．unmerged pull requestsについても分析している．
Another study by Steinmacher~\et~\cite{steinmacher2018ICSE} conducted a survey to also investigate reasons for closed PRs. 
The survey results indicated that the primary reasons for closed PRs were related to duplication and vision mismatches between developers and teams. 
% 準コントリビュータの意識調査をしている．調査の結果，duplicationとvisionの不一致がプルリクエストがマージされない理由に効いてくることが述べられている．また，開発者がabandantしてしまうことも理由の一つだと述べている．

Also, previous studies have investigated the factors influencing the acceptance of PRs~\cite{tsay2014ICSE,iyer2021TSE,rastogi2016ICSE,rastogi2018ESEM,kononenko2018ICSE,terrell2017}.
Tsay~\et~\cite{tsay2014ICSE} reported that acceptance of PRs is influenced by both social and technical factors. 
In particular, the prior interactions of contributors within the project impact the likelihood of acceptance.
% pull requestのアクセプトにはsocialとtechnicalの両面が影響してくることが述べられている．特に，貢献者のプロジェクトにおける事前のやり取りがacceptanceへ影響してくることが述べられている．
Iyer~\et~\cite{iyer2021TSE} investigated the impact of personal traits on the acceptance of PRs. They found that PRs submitted by developers who are more open and conscientious but less extroverted are more likely to be accepted.
% personal traitsを使ってどれくらいPRがアクセプトされるかを分析している．元となっている先行研究にpersonal traitsのメトリクスを追加したという位置付け．
Rastogi~\et~\cite{rastogi2016ICSE,rastogi2018ESEM} investigated the impact of geographical locations of contributors on the acceptance of PRs. They found that geographical location significantly influences PR acceptance.
Also, their study revealed that PRs are more likely to be accepted when contributors and maintainers who review the contributions are from the same country.
% 地理的な位置関係がPRのアクセプトに影響を与えるという論文
% https://ieeexplore.ieee.org/document/7883366
% https://dl.acm.org/doi/10.1145/3239235.3240504
Kononeko~\et~\cite{kononenko2018ICSE} and Pinto~\et~\cite{pinto2018CHASE} investigated the impact of employment relationships on the acceptance of PRs. For example, Pinto~\et~\cite{pinto2018CHASE} found that PRs submitted by contributors employed by the company are more likely to be accepted. 
% 雇用関係が関係するという論文
% https://ieeexplore.ieee.org/document/8449243
% https://ieeexplore.ieee.org/document/8445549
Terrell~\et~\cite{terrell2017} examined the impact of gender on the acceptance of PRs and found that PRs submitted by women are more likely to be accepted.

\smallskip
\noindent \textbf{Prediction Models for PR Outcomes.}
%\mahmoud{this can be another subsection about PRs: Prediction models for PRs outcome}
While many studies have investigated the reasons and factors behind PR decisions, some have focused on constructing prediction models to directly assist maintainers in making these decisions.
Dey and Mockus developed a prediction model to determine the acceptance rate of PRs~\cite{dey2020ESEM}.
The model utilizes 17 metrics that represent technical and social factors. 
Their goal is to identify which factors contribute to the prediction performance. 
The model achieves an ROC AUC of 0.94, and they report that 15 out of the 17 metrics are important for the prediction, such as the number of additions, commits and files modified in a PR.
%\mahmoud{It would be interesting to list the most significant metrics, e.g., top 2 or 3 factors.}
% この研究では，TechnicalおよびSocial factorsを17のメトリクスとして定義し，それらを用いてpull requestのアクセプト率を予測するモデルを作成している．そして，どのメトリクスが予測に寄与するかを明らかにすることを目的としている．3,349のNPM packagesから収集された470,925 PRs（79,128 GitHubユーザが作成）を分析した結果，ROCのAUCが0.94になり，15/17のメトリクスが予測において重要であることが明らかになった．
% Sienくんの研究に近い？
% DONE 読んだ
Fan~\et~\cite{fan2018EMSE} proposed a prediction model to anticipate the acceptance of PRs. They extracted 34 features from code changes in three OSS projects: Eclipse, LibreOffice, and OpenStack. Their proposed model demonstrated a significant improvement in prediction performance. Specifically, their apporach outperformed random guess and two previous approaches~\cite{jeong2009,gousios2014ICSE} by 69\%, 64\%, and 38\%, respectively.
%\mahmoud{what kind of improvements? and improvements with respect to what baseline? by how much this improvement got?}.
% この研究ではあるコード変更がマージされるかどうかを早期に予測するモデルを提案している．OSSにおいてコード変更は検査に時間がかかり，マージされる確率が低い（放棄される）質の低いコード変更へ労力を費やさずに済むことにつながる．具体的にはコード変更から34の特徴を抽出し機械学習モデルによって予測モデルを構築する．166,215のコード変更を3つのOSS（Eclipse，LibreOffice，OpenStack）より収集して実験を行い，提案手法は予測精度を向上させることを確認した．また，放棄されるコード変更とマージされるコード変更を区別する重要な特徴量を明らかにした．
% 早めにこのコードがマージされるかがわかればレビューのprioritizeになるからという理由で変更のメトリクスを使った予測モデルを作成している．
% DONE 読んだ
% Jeong~\et~\cite{}
% % この研究では，FirefoxとMozilla Coreの2つのOSSにおいてbug-fix patchがアクセプトされるかどうかを予測する上で利用できるa set of featuresを提案した．
% % http://rosaec.snu.ac.kr/publish/2009/techmemo/ROSAEC-2009-006.pdf

PRs represent code-level changes (\ie, maintainers and contributors make decisions about code changes). 
In contrast, the proposals examined in this study focus on \emph{design-level decisions}. However, such decisions are often not documented in repositories, resulting in a lack of research on this aspect of development.
% While pull requests represent code-level design decisions, the proposals targeted in this study correspond to design-level decisions.
% %Therefore, it is crucial to conduct an in-depth investigation into these design-level decisions.
% However, such decisions are often not recorded in repositories and few studies have investigated them.
Kruchten~\et~\cite{kruchten2009} emphasized the importance of documenting design decisions and described a set of attributes for architectural design decisions, such as a short textual description of the design decision \emph{Epitome}, a textual description of the reason of the decision \emph{Rationale}, and features where a decision evolves \emph{State}.
%\mahmoud{can we explain/elaborate a bit about this? like what is Epitome?}.
%They also categorized architectural design decisions into four types, such as existing decisions.
%\mahmoud{I didn't understand this sentence :( Can we rephrase?}.

Miesbauer and Weinreich~\cite{miesbauer2013} studied how architectural design decisions are made and documented. They found that the reasons for decisions are often not recorded during the decision-making process. They also emphasized the importance of documenting decisions to exclude certain elements, known as nonexisting decisions, as these are not visible in the code. However, they observed that such decisions are rarely documented.
{I tried to rewrite the part of the text below. Check it please if it is correct and easy to understand. Just tried to simplify the concept of nonexisting decisions.}

\emph{Our study differs from previous work since we specifically focus on declined proposals in the Go programming language, which represent design-level decisions. These proposals are distinct because they address decisions that are often not visible in the code and are rarely documented in repositories.
 Our study investigates the reasons behind these declined proposals, providing insights into the nature and documentation of design-level decisions. Also, we develop a prediction model to support decision-making by contributors and maintainers.
 % \mahmoud{I rewrote the text below in two parts: the scope of our study, and then its goal. I feel the main difference is that our study is about declined proposals (design-decision}
}

% \emph{Our study differs from previous work in two key aspects: (1) our study specifically focuses on declined proposals in the Go programming language, which represent design-level decisions, and (2) it investigates the reasons behind these declined proposals—nonexisting decisions that are often not documented in repositories. By concentrating on declined proposals in the Go programming language, our study aims to provide new insights into the nature and documentation of design-level decisions, as well as to develop a prediction model to support decision-making by contributors and maintainers.}
% 本研究ではdesign decisionsという観点で，declineされるdecisionの理由を調査したところに新規制がある．Miesbauer and Weinreichが主張するように，設計レベルのnonexisting decisionsについての調査はこれまでほとんど行われてこなかった．本研究により，設計レベルのdecisionに焦点を当てた新たな知見を提供することができる．

% \masa{check papers/predicting accepted prs/issues for the references here. Also check the related work in dey2020ESEM. It describes the details of the references}
% % =============================
% % review patterns
% % =============================
% Rigby~\et~\cite{rigby}
% % この研究ではGCC，Linux，Mozilla，Apachの4つのOSSにおけるコードレビュープロセスを分析している．そして，いくつかのレビューパターンを明らかにした．

% related workに書くべきストーリについて考察
% これまでにもOSSに対する貢献がアクセプトされるかどうかを予測する研究は行われている．しかし，これらの研究はアクセプトされる貢献に主眼を置いている．declineされる貢献について深い考察をした研究は少ない．
% 本研究ではGo言語の提案に焦点を当てて，declined貢献について詳しく分析する．
% 新規性？
% li2022TSEやgousios2014ICSE，ではマージされない理由としてtechnical issueよりも文化的な側面が効いてくるおkとが述べられている．それと比較すると，goのproposalsはかなり異なる．

\section{Threats to Validity}
\label{sec:threats}
In this section, we discuss the threats to the validity of our study. 

\subsection{External Validity}
\label{sec:threats:external}
In this paper, we investigated design-level decisions by focusing on proposals from the Go project. While these proposals include many design-level decisions, they may not represent all such decisions in software development. Future studies should examine additional software projects to generalize our findings.

\subsection{Construct Validity}
\label{sec:threats:construct}
We detected initial decision comments using keyword matching with six selected keywords. This approach might have missed some decision comments that do not include these keywords. However, we found that 735 out of 785 proposals (93.6\%) contained at least one of the six keywords, suggesting that our approach covers a large portion of the decision comments. Nonetheless, future studies should explore alternative methods to capture a broader set of decision comments.

We identified resubmitted proposals by using issue numbers to establish traceability links between declined proposals and related issues. Some resubmitted proposals may not include the previous issue numbers, so we might have missed them. Future work should develop alternative approaches to more accurately detect resubmitted proposals.

Furthermore, we manually analyzed the declined proposals to identify the reasons for their decline. This manual analysis may be subjective; to mitigate this threat, at least two authors with extensive programming experience independently analyzed the proposals and discussed the reasons for decline.

\subsection{Internal Validity}
\label{sec:threats:internal}
We used GPT-3.5-turbo to predict declined proposals. OpenAI, the provider of these models, does not disclose the exact model architectures or training data. This lack of transparency may affect the validity of our study, as changes in architecture or training data could alter our results. A possible future research direction is to experiment with other models-—perhaps smaller language models that can be trained on our own servers and datasets-—to assess the robustness of our findings.

\section{Conclusion}
\label{sec:conclusion}

% inferenceとindependentに対する予測モデルを作る必要がある．input dataも考える必要がある．
Design-level decisions in OSS projects are critical for maintainers because once such decisions are made, the software experiences significant changes that must be maintained over a long period. In this paper, we focused on these design-level decisions and examined their characteristics. Specifically, we analyzed proposals from the Go project and evaluated a prediction model for declined proposals along with the reasons behind their decline. We found that 40\% of the proposals were declined and only 14.7\% of the declined proposals were resubmitted. Additionally, we identified nine distinct reasons for the decline of proposals. We then evaluated GPT-3.5-turbo for predicting declined proposals. The results showed that GPT-3.5-turbo achieved an F1 score of 0.88 for predicting declined proposals.
%and could accurately anticipate several reasons for the decline, such as \emph{Duplication} and \emph{Existing alternatives}. Furthermore, GPT-4o improved the performance of model in predicting the reasons behind the declines.
However, we observed that the substantial performance gap between using all comments and using only the proposal title and description. This observation suggests that, while the theoretical upper bound on model performance is high, actual performance in practice remains limited. We found that incorporating the first 25\% of comments provides a reasonable trade-off between cost and benefit.

The following are the main takeaways from this study. 

% GPTをdeclineされるproposalの予測およびその理由を予測することに使うと良い．GPTを使った予測モデルを開発に組み込むことで，declineされるproposalを予測し，特に既存データで解決できる問題についてはdecline理由を事前に明らかにできる．これにより，コントリビュータは早い段階で自分のproposalの改善点がわかり，より良いproposalを再投稿できる．maintainerは議論の時間を短縮できる．

\smallskip{}
\noindent{}
%\mahmoud{Since we have a section for implications, we may not need the list below : )}
%\masa{Sometimes reviewers ask us to add a summary in the conclusion. so let me keep this :)}
\textbf{Recommendation 1: Using GPT-3.5-turbo for predicting declined proposals can be beneficial.}
GPT-3.5-turbo is capable of predicting declined proposals. Such a prediction model could help contributors identify areas for improvement at an early stage, allowing them to revise and resubmit their proposals with a higher chance of acceptance. Also, maintainers could reduce the time and effort spent on discussing proposals that are likely to be declined.

\textbf{Recommendation 2: Using the first 25\% of comments in the proposal discussion is a reasonable trade-off between cost and benefit.}
The prediction model shows improved performance when using discussion comments, but the model does not reliably leverage additional comments. Therefore, we recommend using the first 25\% of comments, as this provides performance gains while minimising contributor waiting time.
We also recommend researchers to propose a new robust prediction model that can handle the comments more effectively. 

% decline判定をする際に，本研究で明らかにした9つの分類を使うと良い．proposalはdeclineされることが多い．本研究で作成した9つのdecline理由は，declineされたproposalの改善理由を探す上で役立つ．declineされたproposalに対してメンテナがその理由を提供すれば，コントリビュータは改善の方向性がわかる．その結果，reappeared proposalsのaccept率が向上する．
\smallskip{}
\noindent{}
\textbf{Recommendation 3: When making a decision on a proposal, we recommend utilizing the list of reasons for declined proposals.}
%As shown in \fig{fig:proposal-statuses}, proposals are often declined.
The nine reasons we identified in this paper can help contributors identify areas for improvement. If maintainers select the appropriate reason from this list, contributors can more easily understand how to improve their proposals. This, in turn, could lead to a higher acceptance rate for resubmitted proposals.

\subsection*{Acknowledgments}
We gratefully acknowledge the financial support of: (1) JSPS for the KAKENHI grants (JP24K02921, JP25K03100, JP25K22845); (2) Japan Science and Technology Agency (JST) as part of Adopting Sustainable Partnerships for Innovative Research Ecosystem (ASPIRE), Grant Number JPMJAP2415, (3) the Inamori Research Institute for Science for supporting Yasutaka Kamei via the InaRIS Fellowship, and (4) the Natural Sciences and Engineering Research Council (NSERC) of Canada.

\bibliographystyle{IEEEtranS}
\balance
\bibliography{reference}

% % Appendix
% \appendix
% \include{sections/appendix}

\end{document}